\documentclass[a4paper, amsfonts, amssymb, amsmath, reprint, showkeys, nofootinbib, longbibliography, twoside]{revtex4-2}
\usepackage[english]{babel}
\usepackage[utf8]{inputenc}
\usepackage[colorinlistoftodos, color=green!40, prependcaption]{todonotes}
\usepackage{amsthm}
\usepackage{mathtools}
\usepackage{physics}
\usepackage{xcolor}
\usepackage{graphicx}
\usepackage[left=23mm,right=13mm,top=35mm,columnsep=15pt]{geometry} 
\usepackage{adjustbox}
\usepackage{placeins}
\usepackage[T1]{fontenc}
\usepackage{lipsum}
\usepackage{csquotes}

\usepackage{amsmath}
\usepackage{setspace}
\usepackage{array} 
\usepackage{amsmath} 

\usepackage{setspace}
\usepackage{parskip}
\usepackage{bm}
\usepackage{color}
\usepackage{physics}
\usepackage{titlesec}
\usepackage{etoolbox}

\definecolor{byzantine}{rgb}{0.74, 0.2, 0.64}

\newcommand{\ve}[1]{{\bf #1}}
\newcommand{\vphi}{\varphi}
\let\vec=\mathbf

\usepackage[pdftex, pdftitle={Article}, pdfauthor={Author}, breaklinks]{hyperref} 
\usepackage{amsmath}
\usepackage{physics}
\usepackage{setspace}
\usepackage{array} 
\usepackage{amsmath} 
\usepackage{graphicx}
\usepackage{setspace}
\usepackage{parskip}
\usepackage{bm}
\usepackage{xcolor}
\usepackage{ulem}

\usepackage{titlesec}
\usepackage{etoolbox}

\let\vec=\mathbf

\begin{document}
\title{
Fast simulation of light scattering and harmonic generation in axially symmetric structures in COMSOL
}

\author{Sergei Gladyshev}
\altaffiliation{School of Physics and Engineering, ITMO University, 191002 St. Petersburg, Russia}

\author{Olesia Pashina}
\altaffiliation{School of Physics and Engineering, ITMO University, 191002 St. Petersburg, Russia}

\author{Alexey Proskurin}
\altaffiliation{School of Physics and Engineering, ITMO University, 191002 St. Petersburg, Russia}

\author{Anna Nikolaeva}

\author{Zarina Sadrieva}
\altaffiliation{School of Physics and Engineering, ITMO University, 191002 St. Petersburg, Russia}

\author{Andrey Bogdanov }
\altaffiliation{School of Physics and Engineering, ITMO University, 191002 St. Petersburg, Russia}
\altaffiliation{Qingdao Innovation and Development Base of Harbin Engineering University, 266400 Qingdao, China}
\email{a.bogdanov@metalab.ifmo.ru}

\author{Mihail Petrov}
\altaffiliation{School of Physics and Engineering, ITMO University, 191002 St. Petersburg, Russia}
\email{m.petrov@metalab.ifmo.ru}

\author{Kristina Frizyuk}
\altaffiliation{School of Physics and Engineering, ITMO University, 191002 St. Petersburg, Russia}
\email{k.frizyuk@metalab.ifmo.ru}

\date{\today} 

\begin{abstract}

In the field of optics and nanophotonics, simulation of electromagnetic scattering plays a major role in the study of complex nanostructures and optical devices. The numerical analysis of scattering spectra, even for nanocavities with simple geometry, is associated with significant computational difficulties. However, when the system exhibits certain symmetries, it becomes possible to simplify the problem through the process of separation of variables, which leads to a decrease in its dimension. 
In this paper, we aim to provide a practical guide to a fast simulation of linear and non-linear scattering problems in COMSOL Multiphysics\textsuperscript{\textregistered} for axisymmetric objects including computation of scattering cross-section as well as its multipolar decomposition, optical forces, and second harmonic generation. We also accompany the provided guide with the ready-to-run COMSOL\textsuperscript{\textregistered} models.    

\end{abstract}


\maketitle

\section{Introduction} \label{sec:intro}

Numerical simulations play a crucial role in optics and nanophotonics since they can describe the optical properties of complex nanostructures and devices without their fabrication and direct experimental characterization. 
Numerical optimization became an integral part of the research pipeline~\cite{rumpf2022electromagnetic,lavrinenko2018numerical,lavrinenko2004comprehensive,gallinet2015numerical}  improving the performance of optical devices.
The modern methods of computational electrodynamics allow one to analyze the interaction of light with complex optical systems accounting for a nonlocal and nonlinear response~\cite{schmitt2016dgtd,mortensen2014generalized,fan2006second,kippenberg2018dissipative,itina2002laser}, molecular dynamical~\cite{zeng2016recent,benz2016single} and quantum mechanical effects~\cite{zhu2016quantum,yang2019general,christensen2017quantum}.  
 
The full-wave numerical simulation of real experimental samples or optical devices is time-consumable and requires essential computational facilities.
A detailed analysis of linear and nonlinear scattering spectra of scatterers (nanoresonators or metaatoms) is crucial for designing the nanophotonic system functionality.
Even a simple scattering task can be quite challenging in terms of computational resources when it comes to optimization problems~\cite{Wiecha2017Feb} or for inverse design of nanophotonic systems~\cite{Liu2018Oct}.
However, if the scattering potential has specific symmetries, the scattering (or eigenvalue) problem can be essentially simplified via the separation of variables and the effectively reducing the dimensionality of the problem.
After that, the reduced problem can be solved numerically much faster than the initial one.
This approach is universal and can be combined with various numerical methods like the finite-element method (FEM)~\cite{KOSHIBA2014}, finite-difference methods~\cite{Lusse1994,Hadley1995,SanginKim1996}, method of moments~\cite{Jakobus1995}, or others~\cite{gallinet2015numerical, Scali2023-Graphtheoryapproach,Vavilin2023-ThePolychromaticT-m}.  

The case of scatterers of cylindrical symmetry gains a lot of interest due to their relatively simple methods of their fabrication with modern methods of nanotechnology, and, at the same time, they are ideal elementary blocks of complex nanophotonic systems.
Cylindrical scatterers have already demonstrated a wide range of nanophotonic effects such as resonant Kerker effect~\cite{Geffrin2012Nov, Alaee2015-AgeneralizedKerker}, perfect absorption~\cite{Proskurin2021Aug}, and achieving high-Q resonant states in single structures~\cite{Pichugin2023-Aseriesofavoidedc, Koshelev2020-Subwavelengthdielect, Khademalrasool2021-Rapidsynthesisofsi, deCeglia2019-Second-HarmonicGener, Chaliyawala2019-Effectivelightpolar, Mignuzzi2019-NanoscaleDesignoft, Arumona2023-Material-andshape-d}.
For the scatterers with cylindrical symmetry, the separation of the azimuthal variable allows reducing the problem dimension from three-dimensional (3D) to two-dimensional (2D).
Then the reduced 2D problem can be solved numerically.
This approach is widely used for the calculation of light scattering from rotationally symmetric particles via T-matrix methods~\cite{mishchenko1996t,mishchenko1998capabilities}.
A special interest is to implement such a method in modern numerical simulation software such as COMSOL Multiphysics\textsuperscript{\textregistered}.
In Ref.~\onlinecite{oxborrow2007traceable}, Mark Oxborrow firstly implemented the rotational symmetry approach in COMSOL\textsuperscript{\textregistered} for finding the spectra of whispering gallery modes in resonators of various shapes.
Later, 2D axial symmetry module was built in COMSOL\textsuperscript{\textregistered}~\cite{ BibEntry2023Jul61} as a default setting. It may seem that this module does not allow to solve the scattering problem under the arbitrary angle of incidence as obliquely incident waves break the rotational symmetry of the problem. Nevertheless,  the obliquely incident wave can be expanded into a Fourier series over $e^{-im\varphi}$ and then the scattering for each harmonic can be calculated independently. 
This efficient approach was implemented in COMSOL for scalar acoustic field~\cite{2dopticsCOMSOL3} and recently for electromagnetic waves~\cite{2dopticsCOMSOL, 2dopticsCOMSOL3, 2dopticsCOMSOL2}. 

In this work, we have gone far beyond and  provide a comprehensive guide:
(i) on how to efficiently solve both linear and nonlinear (second-harmonic generation) electromagnetic scattering taking advantage of the rotational symmetry of the scatterers;
(ii) calculate the scattering cross-section, its multipolar decomposition in 2D axisymmetric systems;
(iii) calculate the Maxwell stress tensor and the Cartesian components of the optical force in the cylindrical basis.
Though the proposed approach is universal and can be realized in various numerical packages, we have applied it for COMSOL Multiphysics\textsuperscript{\textregistered}  as it is one of the most spread tools for electromagnetic simulations.
Moreover, our method does not require the additional built-in features, and possible for realization starting at least from version 5.5.
We have already successfully used it for simulating the optical properties of resonant nanoantennas on a substrate~\cite{Sinev2016Sep}, excitation of surface plasmon polaritons by spherical and cylindrical nanoantennas~\cite{Sinev2020Mar,Dvoretckaia2020Mar}, calculation of harmonic generation in resonators with rotational symmetry~\cite{Toftul2023-Nonlinearity-Induced}, optical forces acting on particles above structured substrates~\cite{Ivinskaya2018Nov,Kostina2020Feb}, and perfectly absorbing nanoantennas on a conducting surface~\cite{Proskurin2021Aug}.
While the suggested approach is used for certain tasks in the mentioned papers, they don't contain a detailed technical description of the calculation methods. Here we fill this gap and provide a comprehensive practical guide to solving linear and nonlinear scattering problems for systems with rotational symmetry in COMSOL Multiphysics\textsuperscript{\textregistered}.

\section{Axial symmetry from 3D to 2D} \label{sec:2dgeometry}

\begin{figure}[t]
\begin{center}
    \includegraphics[width=0.5\textwidth]{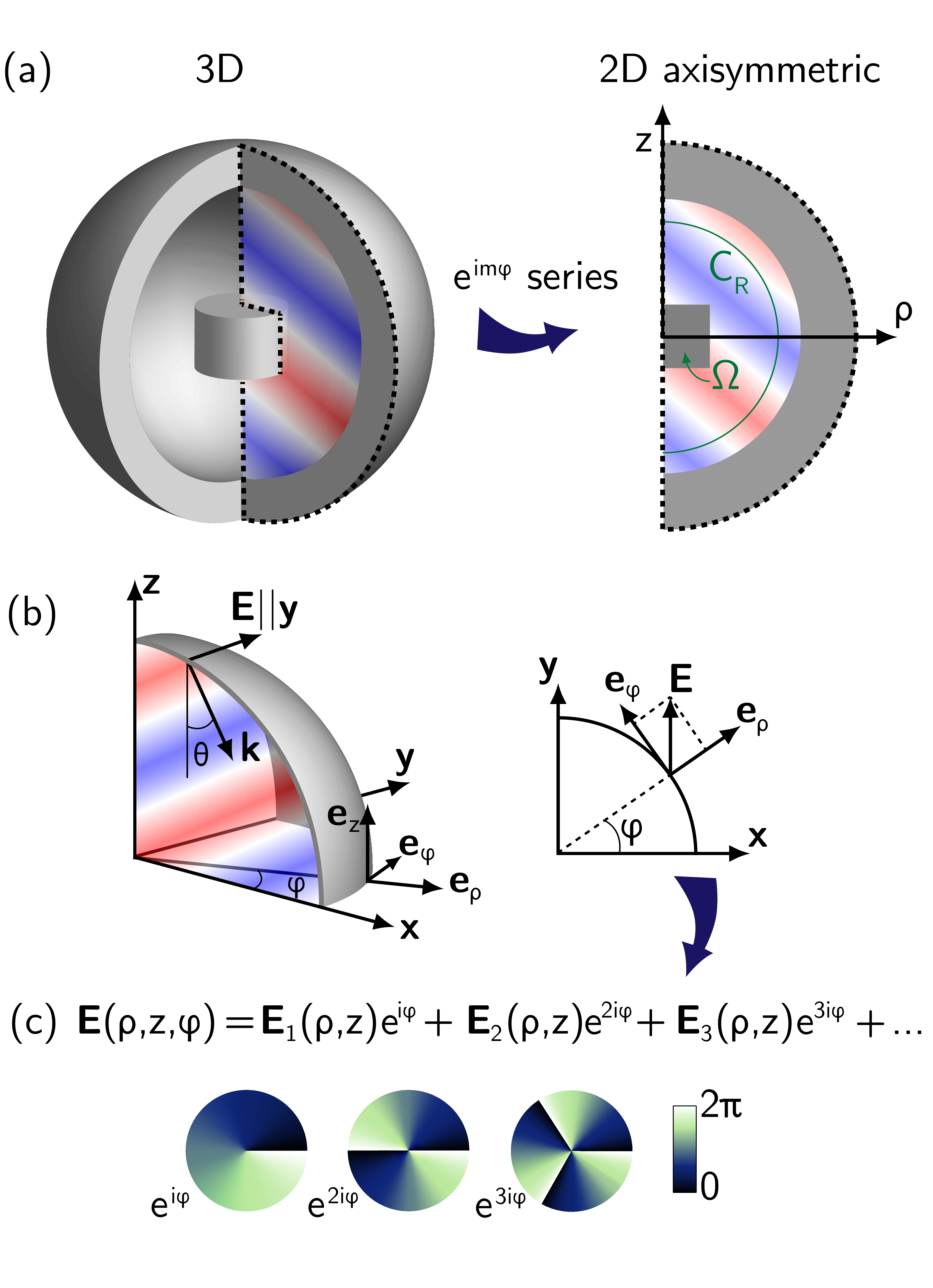}
    \caption{Moving from 3D to 2D. Calculation of electromagnetic properties of the system in the 2D model. As the example,  it is shown in detail how the components of the electromagnetic field can be rewritten into a series $\mathbf{E}_{m}(\rho,z) e^{-i m \varphi}$ for TE polarization.}
    \label{fig:concept}
\end{center}
\end{figure}

The method bases on the reduction of 3D problem to 2D problem by the expansion of the electromagnetic fields $\mathbf{E}(\mathbf{r})$ into a Fourier series of waves corresponding to different azimuthal indices $m$~\cite{Chirikjian2000Sep}:
\begin{equation}
\label{eq:cyl_decomp}
    \mathbf{E}(\mathbf{r}) = \sum_{m = -\infty}^{\infty} \mathbf{E}_{m}(\rho, z) e^{-i m \varphi}. 
\end{equation}
Here, $\vb{E}_{m}(\rho, z)$ represents the field components in cylindrical coordinates ($\rho$, $\varphi$, $z$), and $m$ is the number associated with the respective azimuthal harmonic (see Fig.~\ref{fig:concept}). 
The total field $\mathbf{E}(\mathbf{r})$ can be represented as a sum of the incident (background) $\mathbf{E}^{\text{inc}}(\mathbf{r})$ and scattered $\mathbf{E}^{\text{scat}}(\mathbf{r})$ fields
\begin{equation}
   \vb E (\rho, \varphi, z) =  \vb E^{\text{inc}} (\rho, \varphi, z)+\vb E^{\text{scat}} (\rho, \varphi, z). 
\end{equation}
This formalism is implemented in COMSOL Multiphysics\textsuperscript{\textregistered}~\cite{BibEntry2017Apr}.
Its advantage in the accuracy of calculation becomes crucial when the magnitude of the scattered field is much smaller than one of the incident field.
Both incident and scattered fields can be expanded into a Fourier series:
\begin{equation}
    \label{eq:Scat_tot}
    \vb E^{v} (\rho, \varphi, z) = 
    \sum_{m = -\infty}^{\infty} \vb E_m^{v} (\rho,z) e^{-i m \vphi},
\end{equation}
where $v=\{\text{inc, scat}\}$.
In virtue of the axial symmetry of the problem, a Fourier amplitude of the incident field $\vb E_m^\text{inc} (\rho,z)$ induces only the Fourier amplitude of the scattered field $\vb E_{m'}^\text{scat} (\rho,z)$ with the same azimuthal index, i.e. $m = m'$.
One can say that the azimuthal harmonics with different indices $m$ do not mix with each other~\cite{Noether1918-InvarianteVariations, Gladyshev0, Xiong2020-Ontheconstraintsof}.
Therefore, each Fourier amplitude of the scattered field $\vb E_{m'}^\text{scat} (\rho,z)$ for each $m$ can be calculated independently.
Then, taking a sum over $m$ [see Eq.~\eqref{eq:Scat_tot}] one retrieve the scattered field in 3D space.

    Formally, the Fourier expansion \eqref{eq:cyl_decomp} reduces a 3D problem to an infinite set of 2D problems as the series is infinite.
However, if the maximal radial size $R_\text{max}$ of a scatterer is not large,
$R_\text{max}k_0\sin\theta\lesssim 1$,
the Fourier series \eqref{eq:cyl_decomp} converges fast and only a few terms is enough to describe the scattered field accurately.
Here $k_0$ is the wavenumber of the incident plane wave in the surrounding space, $\theta$ is the angle of incidence.
Thus, $m \in [-M_{\text{max}} .. M_{\text{max}}]$, where the truncation number $M_{\text{max}}$ can be estimated from the empiric rule as 
$M_{\text{max}}\approx R_\text{max}k_0\sin\theta$.
A more accurate analysis of the truncation number and its connection with precision can be found in~\cite{Rohfritsch2019Jun,Song2001Jul,Mishchenko2002Jun}.

Therefore, the problem of linear scattering from the axially symmetric structure of a finite size can be reduced to a finite number of 2D scattering problems.
It is also worth mentioning that due to orthogonality of azimuthal functions $e^{-im\varphi}$ with different $m$, they correspond to independent scattering channels.
Thus, the Fourier expansion \eqref{eq:cyl_decomp} not only allows for accelerating the calculations but also gives important physical information on how the scattered power redistributed over the scattering channels.
Below we provide hands-on formulas for scattering, extinction, absorption cross-sections, Maxwell stress tensor, and optical force in terms of 2D harmonics.

\subsection{Scattering cross-section}
The Poynting vector for the scattered field corresponding the angular harmonic $e^{-im\varphi}$ can be written as 
\begin{equation}\label{eq:9}
    \vb S_m^\text{scat} = \dfrac 12 \text{Re}[\vb E_m^\text{scat} \times\vb H_m^\text{scat*}], 
\end{equation}
Thus, the partial scattering cross-section is 
 \begin{equation}
 \begin{aligned}
    \sigma_m^\text{scat} = & \frac{1}{I_{\text{inc}}}\int_{S^2} (\vb S_m^\text{scat}\cdot \vb n)\ \dd s = \\ = &
    \frac{1}{I_{\text{inc}}}\int_{C_R} 2\pi \rho (\vb S_m^\text{scat}\cdot \vb n)\ \dd c,
\end{aligned}
\end{equation}
where $I_{\text{inc}}= |E^{\text{inc}}|^2/(2Z)$ is the energy flux of the incident wave,  $Z = \sqrt{{\mu_0}/{\varepsilon_0}}$ is the impedance of the embedding medium (vacuum in our case), and  the integral is taken over the sphere $S^2$ surrounding the structure. 
In 2D axial symmetry this sphere become semi-circle $C_R$ (see Fig.~\ref{fig:concept}).
For 2D geometry integral over the angle $\varphi$ gives the multiplier 
$2\pi\rho$, and  $\dd c$ is the circle arc length differential.
We note that COMSOL\textsuperscript{\textregistered} allows omission of the $2\pi\rho$ multiplier if the \texttt{Compute surface integral} option is selected.
See details of the derivation in the Supplemental Material.

Due to orthogonality of the electromagnetic modes with different $m$, the total scattering cross-section can be obtained by summing over all orders $m$: 
 \begin{equation}\label{eq:10}
    \sigma^{\text{scat}}=\sum_{m = 0}^{\infty} (2-\delta_{0,m}) \sigma_m^\text{scat} ,
 \end{equation}
where $\delta_{0,m}$ is the Kronecker symbol, which appears due to $\sigma_m^\text{scat} = \sigma_{-m}^\text{scat}$, according to the properties of the problem {[see Eq.~\eqref{eq:even-odd}]}.

\subsection{Absorption cross-section}

As well as scattering cross-section, the total absorption cross-section $\sigma^\text{abs}$ can be calculated as a sum of partial absorption cross-sections $\sigma^\text{abs}_m$:
\begin{equation}
\label{eq:t_abs_CS}
    \sigma^\text{abs} = \sum_{m = 0}^{\infty}
    (2 - \delta_{0,m})\sigma^\text{abs}_m.
\end{equation}
Each partial absorption cross-section can be calculated through the following volume integral:
\begin{equation}
\label{eq:p_abs_CS}
    \sigma^\text{abs}_m=\frac{\omega\pi}{I_\text{inc}}\iint_\Omega
    \text{Im}\left\{\mathbf{P}_m^*\mathbf{E}_m\right\} \rho \mathrm{d}\rho\mathrm{d}z.
\end{equation}
Here, $\mathbf{P}_m=\varepsilon_0(\varepsilon-1)\mathbf{E}_m$ is the polarization, $\varepsilon$ is the dielectric permittivity of the scatterer's material.
The integral is taken over the cross-section area $\Omega$ of the scatterer [see Fig.~\ref{fig:concept}(a)].
See details of the derivation in the Supplemental Material.

\subsection{Extinction cross-section}

The total extinction cross-section $\sigma^\text{ext}$ can be calculated as the sum of the partial extinction cross-sections $\sigma^\text{ext}_m$ corresponding to different $m$ by analogy with Eqs.~\eqref{eq:10} and \eqref{eq:t_abs_CS}.
The partial extinction cross-sections can be calculated in several ways:
\begin{itemize}
    \item[(i)] By definition of the extinction cross-section:
    \begin{equation}
        \sigma^\text{ext}_m=\sigma^\text{scat}_m+\sigma^\text{abs}_m.
    \end{equation}
    \item[(ii)] By taking a surface integral over the cross-section area $\Omega$ of the scatterer [see Fig.~\ref{fig:concept}(a)]:
    \begin{equation}
    \label{eq:p_ext_CS}
        \sigma^\text{ext}_m=\frac{\omega\pi}{I_\text{inc}}\iint_\Omega
        \text{Im}\left\{\mathbf{P}_m^*\mathbf{E}^\text{inc}_m\right\} \rho \mathrm{d}\rho\mathrm{d}z.
    \end{equation}
    \item[(iii)] {By taking a line integral over $C_R$ [see Fig.~\ref{fig:concept}(a)]:
    \begin{equation}
        \sigma_m^\text{ext} = \frac{1}{I_{\text{inc}}}\int_{C_R} 2\pi  \rho (\vb S_m^\text{ext}\cdot \vb n)\ \dd c,
    \end{equation}
    where
    \begin{equation}
        \vb S_m^\text{ext}=-\frac{1}{2}\text{Re}
        \left\{
        \mathbf{E}_m^\text{inc}\times\mathbf{H}_m^{\text{scat}*}+
        \mathbf{E}_m^\text{scat}\times\mathbf{H}_m^{\text{inc}*}
        \right\}.    
    \end{equation}}
\end{itemize}
Therefore, scattering, absorption and extinction cross sections of an axially symmetric scatterer can be calculated using both surface or line integral.
See details of the derivation in the Supplemental Material.

\subsection{Maxwell stress tensor and optical forces}

Optical force is widely studied in nanooptics and nanophotonics as it allows for trapping and manipulating micro- and  nanoobjects via optical fields~\cite{Ashkin1997May,Marago2013Nov,Ivinskaya2017May}.    While the optical force can be directly computed by integrating the Maxwell stress-tensor over the area containing the scatterer, one still needs to know the electromagnetic fields distribution in the near- or far-zones.
One can connect the terms with different $m$ and optical forces in order to effectively compute the forces acting on a scatterer with rotational symmetry.
Indeed, the Maxwell stress-tensor has the form \cite{Novotny2012Sep}
\begin{equation}
\label{eq:Max_st_tens}
    \widehat{T} = \varepsilon_{0}\mathbf{E} \otimes \mathbf{E} + \mu_{0}\mathbf{H} \otimes \mathbf{H} - \frac{1}{2}\left( \varepsilon_{0}\mathbf{E}\mathbf{E} + \mu_{0}\mathbf{H}\mathbf{H} \right)\widehat{I}.
\end{equation}
Here $\varepsilon_{0}$ and $\mu_{0}$ are the permittivity and permeability of vacuum, $\widehat{I}$ is the identity tensor.
The optical force can be calculated by integration of Eq.~\eqref{eq:Max_st_tens} over the closed area containing the scatterer: 
\begin{align}
\label{eq:force}
    \vb F=\oint_{S^2}\widehat{T} \mathrm{d}s. 
\end{align}
Substituting expansion~\eqref{eq:cyl_decomp} into Eq.~\eqref{eq:force} one can reduce the integration to the integration over the line $C_R$ and summation over the harmonics.
The $x$-component of the force acting on the scatterer has the following form
 \begin{equation}
     \left\langle F_x\right\rangle = \left\langle F^\text{E}_x\right\rangle + \left\langle F^\text{H}_x\right\rangle,
 \end{equation}
where
\begin{equation}
\begin{aligned}
    \left\langle {F}^\text{E}_{x}\right\rangle 
    & = \frac{\varepsilon_0}{4}\mathop{\mathrm{Re}}\int_{C_R}^{}{2\pi {\rho}\, \mathrm{d}c}\,\sum_{m = - \infty}^{ \infty}\left\lbrack \left( E_{m,\rho}\left( E_{m + 1,\rho} \right)^{*} - \right.  \right.  \\ 
    & -  \left. \left. E_{m,\varphi}\left( E_{m + 1,\varphi} \right)^{*} - E_{m,z}\left( E_{m + 1,z} \right)^{*} -\right.\right.\\
     & - \left. \left. iE_{m,\rho}\left( E_{m + 1,\varphi} \right)^{*} + iE_{m + 1,\rho}\left( E_{m,\varphi} \right)^{*} \right)n_{\rho}+\right.  \\
     & + \left. 2E_{m,\rho}\left( E_{m,z} \right)^{*}n_{z} \right \rbrack.
\end{aligned}
\end{equation}
Here $n_{\rho}$ and $n_z$ are the coordinate-dependent components of the vector normal to the integration surface,  and $\left\langle F^\text{H}_x\right\rangle$ satisfies the same equation after replacing $\varepsilon_0$ with $\mu_0$ and $\vb E$ with $\vb H$.
We refer readers to the Supplementary Materials, where they can find details on the derivation of the formulas above and the expressions for other components of the optical force.

We append to our paper a COMSOL Multiphysics$^{\mbox{\scriptsize\textregistered}}$ file~\cite{gitmultipole} 
that calculates the values of the optical force components for the simplest case of a plane wave incident on a single spherical nanoparticle in a vacuum.
Since the proposed method applies to any axisymmetric system, it is also convenient for more complicated cases.
For example, we used it to investigate optomechanical properties of nanoobjects above the substrates with hyperbolic dispersion~\cite{Ivinskaya2018Nov, Kostina2020Feb}.
\section{Linear scattering of a plane wave} \label{sec:scattering}

\begin{figure}[t]
\begin{center}
    \includegraphics[width=0.49\textwidth]{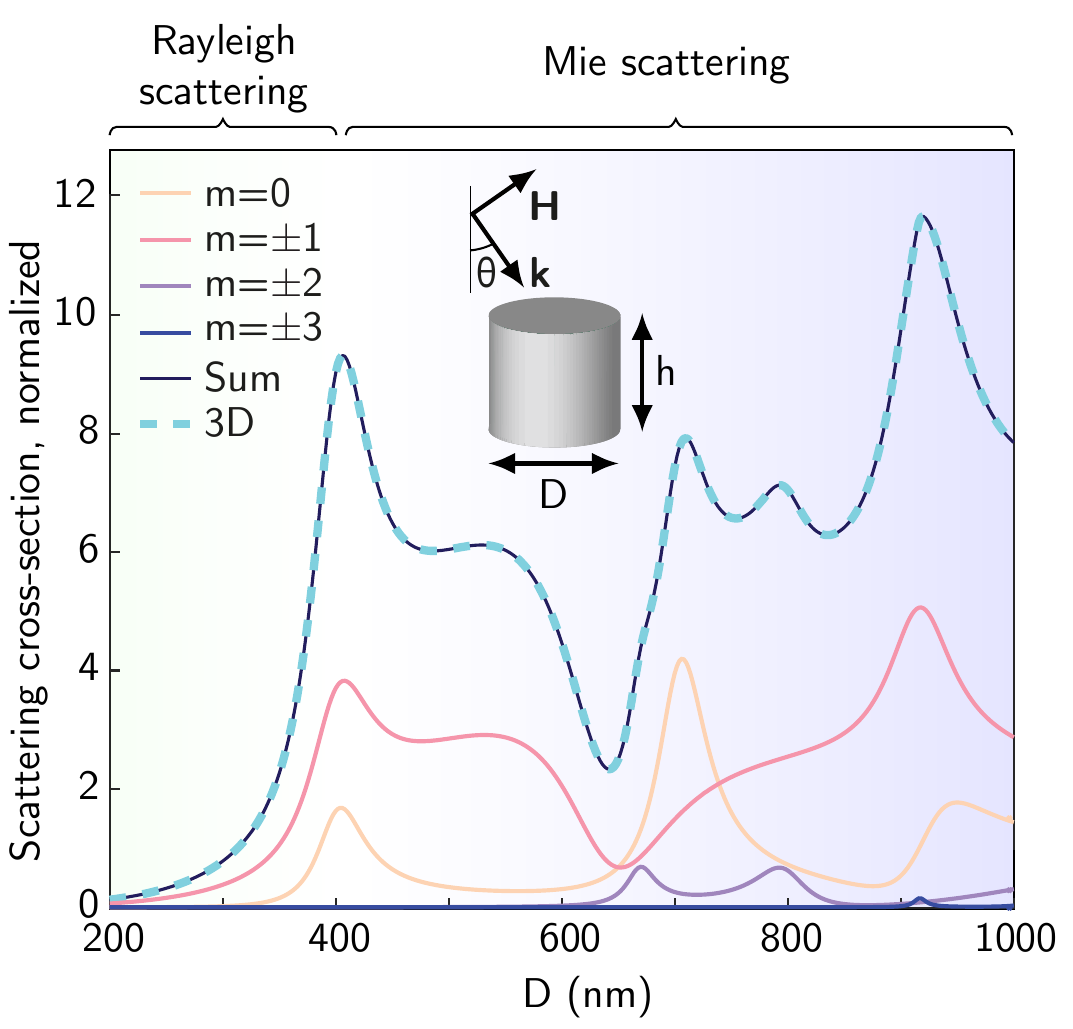}
    \caption{Spectra of the total scattering cross-section $\sigma_{\text{scat}}$ normalized on $h\cdot D$ of the semiconductor  cylinder resonator~(GaAs) with height $h=400$~nm  as a function of diameter $D$ for TE polarization at an angle of incidence of $\theta=30^\circ$. Wavelength of the incident wave $\lambda=1550$~nm.
    }
    \label{fig:scat}
\end{center}
\end{figure}
The formulated approach can be illustrated by an example of a TE-polarized plane wave scattering on a dielectric cylinder.
Let us consider a plane wave incident on the cylinder at an angle $\theta$ [see Fig.~\ref{fig:concept}(b)].
The $\vec k_0$-vector lies in the $xz$-plane, while the $\vec E$-field has only $y$-component. Thus, the wavevector has only two components    
\begin{equation}
\begin{aligned}
    \vec k_0 = k_{0z}\vec e_z + k_{0x} \vec e_{x}. 
\end{aligned}
\end{equation}
The incident electric field in cylindrical coordinates will have the following form: 
\begin{equation}
     \vec {E}^{\text{inc}} =
     \left(
     \begin{array}{c}
        E_{\rho}^{\text{inc}} \\ 
        E_{\varphi}^{\text{inc}}  \\ 
        E_z^{\text{inc}}
    \end{array}
    \right) =
    \left(
    \begin{array}{c}
        {E_{0} \sin \varphi} \\
        {E_{0} \cos \varphi} \\ 
        {0}
    \end{array}
    \right)
    e^{i k_{0z} z-i k_{0x} \rho \cos \varphi}. \\ 
\end{equation}
One can expand the incident field into the series over $e^{-im\varphi}$  using Jacobi-Anger expansion~\cite{Chirikjian2000Sep}:
\begin{equation} 
    e^{-i k_{0x}\rho \cos \varphi} = \sum_{m=-\infty}^{\infty} (-i)^m\, J_m(k_{0x}\rho )\, e^{-i m \varphi}.
\end{equation}
The radial and azimuthal components of the field read 
\begin{equation}\label{eq:rho_te}
\begin{aligned}
    E_{\rho}^{\text{inc}} = \sum_{m = -\infty}^{+\infty} \underbrace{E_{0}e^{i k_{0z}z} (-i)^{m+2} \frac{m}{k_{0x} \rho} J_{m}\left(k_{0x} \rho\right)}_{E_{m,\rho}} e^{-i m \varphi},
\end{aligned}
\end{equation}
\begin{equation}\label{eq:varphi_te}
\begin{aligned}
    E_{\varphi}^{\text{inc}} = \sum_{m=-\infty}^{+\infty} \underbrace{E_{0} e^{i k_{0z}z}(-i)^{m-1} \frac1{k_{0x}}\frac{\dd J_{m}(k_{0x}\rho)}{\dd \rho}}_{E_{m,\varphi}} e^{-i m \varphi}.
\end{aligned}    
\end{equation}
One can notice that the $\varphi$-component of the field is even, while the $\rho$-component is odd
\begin{equation}
\begin{aligned} \label{eq:even-odd}
    E_{m, \varphi} =  E_{-m, \varphi},\quad  E_{m, \rho} = -E_{-m, \rho}.
\end{aligned}    
\end{equation}

The expressions for the case of the TM-polarization are provided in the Supplemental Material.

Once the harmonic amplitudes are found numerically, the scattering cross-section can be computed by using Eqs. \eqref{eq:9}--\eqref{eq:10}.
Here, as an example, we consider a semiconductor cylinder made of GaAs material located in a free space.
The choice of material is provided by the fact that semiconductor materials are widely utilized as material platform for nanophotonics~\cite{Kneissl2020Jan}.
On top of that, GaAs has large second-order nonlinear susceptibility, in particular, $\hat\chi^{(2)}$~\cite{Ghalgaoui2018Dec} responsible  for generation of the second-harmonic, which is discussed in Sec.~\ref{sec:SHG}.   

Figure \ref{fig:scat} shows the partial normalized scattering cross-sections as a function of the cylinder diameter $D$ for $m \in [-3..3]$ calculated in 2D axisymmetric model, and the total cross-section calculated in the 3D model.
The height of the cylinder in the considered case is fixed as $h=400$ nm;
the wavelength of the incident wave is $\lambda=2\pi/k_0=1550$~nm, the angle of incidence $\theta=30^{\circ}$.
Since the harmonic amplitudes decay fast with the harmonic number $m$, the sum of the partial cross-sections sharply converges to
the total cross-section obtained in the 3D simulation.
One can also see the resonant behavior in the scattering spectra which is clearly associated with the excitation of Mie resonances, while at small diameters, the scattering cross-section decreases manifesting the Rayleigh regime of scattering. 
Note that the term with a particular $m$ refers to the sum of all possible vector spherical harmonics~\cite{Stratton} with this angular momentum projection.
Thus, by looking at such an expansion we can only partially extract the multipolar decomposition.
However, since each mode of a cylindrical structure consists of
an infinite sum of multipoles with the same $m$~\cite{Gladyshev0, Xiong2020-Ontheconstraintsof}, one can tell which type of mode is excited.
\section{Multipolar decomposition}
\label{sec:multipole decomposition}
The multipolar decomposition is a powerful tool in electromagnetic scattering theory~\cite{McLean1967Sep,Mitroy2010Oct,Muhlig2011Jun,Yang2017Apr,Ziolkowski2017Jul,Alaee2018-Anelectromagneticmu,Terekhov2019Jan,Gurvitz2019May,Mun2020May,Ray2021Jul}, which allows for predicting the optical response of either compact scatterers or their finite or infinite arrays.
It is based on the idea that any electromagnetic field can be expanded over the series of vector spherical harmonics {(VSHs)}~\cite{Bohren1998Mar, Stratton}.
Despite that the alternative Cartesian multipole decomposition is also found very useful for many particular applications~\cite{Gurvitz2019May,Evlyukhin2016Nov}, we will focus on the  expansion  in this paper.
Accordingly, the scattered electric field is decomposed into multipolar fields in SI units as 
\begin{equation}
\label{eq:VSH_decomp}
   \vb{E}(\vb{r}) =Z \sum_{j=1}^{\infty}\sum_{m=-j}^{j} i a_{jm} \vb{N}^{(3)}_{jm}(\vb{r}) + b_{jm}\vb{M}^{(3)}_{jm}(\vb{r}),
\end{equation}
where $a_{jm}$, $b_{jm}$ are the coefficients characterizing the contribution from the electric $\vb{N}^{(3)}_{jm}(\vb{r})$ and magnetic $\vb{M}^{(3)}_{jm}(\vb{r})$ vector spherical harmonics~\cite{alaee2019exact}, where the radial part $h_n(k_0r)$ is a spherical Hankel function, related to the outgoing wave (see the comparison of VSHs definitions in Suppl. Info of~\cite{Toftul2023-Nonlinearity-Induced}).

In the following part of this section, we provide the link between the azimuthal Fourier 2D expansion [Eq.~\eqref{eq:cyl_decomp}] and spherical multipole decomposition [Eq.~\eqref{eq:VSH_decomp}].  
Specifically, we  show how to make the multipole decomposition, give the exact expressions of the multipolar coefficients, and compare the numerical results from COMSOL$^{\mbox{\scriptsize\textregistered}}$ with the exact results obtained using Mie theory~\cite{BibEntry2023Jul, Bohren1998Mar}. We also accompany our analysis with the ready-to-run COMSOL$^{\mbox{\scriptsize\textregistered}}$ model~\cite{gitmultipole}. 

The expressions for the multipole coefficients in the spherical basis~\cite{alaee2019exact, Alaee2018-Anelectromagneticmu,Fernandez-Corbaton2015-Exactdipolarmoments}:
\begin{equation}\label{eq:99}
\begin{aligned}
     a_{j m}= -(i)^{j-1} \frac{k_0^{2}}{{2 \pi}}
     \sum_{\bar{\ell} \bar{m}}(-i)^{\bar{\ell}} \int \dd {\widehat{\mathbf{p}}} \  \mathbf{Z}_{j m}^{\dagger}(\widehat{\mathbf{p}}) Y_{\bar{\ell} \bar{m}}(\widehat{\mathbf{p}}) \cdot\\
     \cdot \int \dd[3] \mathbf{r} \  \mathbf{J}(\mathbf{r}) Y_{\bar{\ell} m}^{*}(\widehat{\mathbf{r}}) j_{\bar{\ell}}(k_0 r),
 \end{aligned}
 \end{equation}
 \begin{equation}
 \begin{aligned}\label{eq:100}
     b_{j m}=-(i)^{j} \frac{k_0^{2}}{2  \pi}\sum_{\bar{\ell} \bar{m}}(-i)^{\bar{\ell}} \int \dd \widehat{\mathbf{p}} \  \mathbf{X}_{j m}^{\dagger}(\widehat{\mathbf{p}}) Y_{\bar{\ell} \bar{m}}(\widehat{\mathbf{p}})\cdot 
     \\ \cdot \int \dd[3] \mathbf{r} \ \mathbf{J}(\mathbf{r}) Y_{\bar{\ell} m}^{*}(\widehat{\mathbf{r}}) j_{\bar{\ell}}(k_0 r),
  \end{aligned}
\end{equation}
where $\bar{m} \in\{ -\bar{\ell} \dots \bar{\ell}\}$.
For electric component $a_{jm}$, index $\bar{\ell}$ takes only two allowed values $\bar{\ell}\in\{j-1,j+1\}$, while for magnetic multipoles $b_{jm}$, it takes only one allowed value $\bar\ell=j$.
The  $j_\ell(k_0r)$ is the spherical Bessel function.
The $Y_{\ell m}\colon S^2 \to \mathbb C$ is
the scalar spherical harmonics defined as in Ref.~\onlinecite{jackson1999classical}.
The symbol $\widehat{\mathbf{p}}={\vb {p}}/{|\vb p|}$ represents  the  angular  part  of  the  momentum  vector $\mathbf{p}$, where $|\mathbf{p}|={\omega}/{c} $.  
Vector $\widehat{\mathbf{r}}={\vb {r}}/{|\vb r|}$ is the unit vector along $\vb r$.

The $\mathbf{Z}_{j m}(\hat{\mathbf{p}}), \mathbf{X}_{j m}(\hat{\mathbf{p}})$  are the multipolar functions in momentum space defined as
\begin{equation}
    \mathbf{X}_{j m}(\hat{\mathbf{p}})=\frac{1}{\sqrt{j(j+1)}} \mathbf{L} Y_{j m}(\hat{\mathbf{p}}),
\end{equation}
\begin{equation}
    \mathbf{Z}_{j m}(\hat{\mathbf{p}})=i \hat{\mathbf{p}} \times \mathbf{X}_{j m}(\hat{\mathbf{p}}).
\end{equation}

The current density corresponds  to the polarization vector as $\vb {J}(\vb{r}) =
i \omega  \mathbf{P}(\mathbf{r})=i\omega\varepsilon_0(\varepsilon-1)\mathbf{E}$\footnote{It worth mentioning that in COMSOL$^{\mbox{\scriptsize\textregistered}}$, the variable {\fontfamily{qcr}\selectfont J} corresponds to the displacement current but not to the polarization current. Therefore in COMSOL$^{\mbox{\scriptsize\textregistered}}$, {\fontfamily{qcr}\selectfont
J=i$\omega$D
}  but {\fontfamily{qcr}\selectfont
J$\neq$i$\omega$P}. Thus, to avoid mistakes, we recommend paying great attention to the choice of variables in postprocessing and double-checking their definitions in COMSOL$^{\mbox{\scriptsize\textregistered}}$.
}
as the harmonic time dependence in COMSOL$^{\mbox{\scriptsize\textregistered}}$ is defined as $e^{i \omega t}$.
It and can be expanded into a Fourier series
\begin{equation}
    \mathbf{J}(\rho,z,\varphi)=\sum_{m=-\infty}^{\infty} \mathbf{J}_{m}(\rho,z) e^{ - i m \varphi}.
\end{equation}
In the following we will use the components of the current $J_{m\rho}$, $J_{m\varphi}$, $J_{mz}$, while the small $j$ stands for spherical Bessel functions. 

The total power radiated is a sum of contributions from the different multipoles:
\begin{equation}
    P^{\text{scat}} = \frac{Z}{2k_0^2}\sum_{j,m}  \left(|a_{jm}|^2+|b_{jm}|^2\right).
\end{equation}
The scattering cross sections $\sigma^{\text{scat}}$ are defined from $P_{\text{scat}}$ by normalization to the energy flux of the incident wave $I^{\text{inc}} = |E^{\text{inc}}|^2/(2Z) $
\begin{equation}
    \sigma^{\text{scat}} = \frac{Z^2}{k_0^2 |E^{\text{inc}}|^2} \sum_{j,m}  \left(|a_{jm}|^2+|b_{jm}|^2\right).
\end{equation}

We provide a set of scripts~\cite{gitwolfram} that help in the computation of exact multipolar moments for systems with axial symmetry.
The scripts provide the expressions for the coefficients $a_{j m}$, $b_{j m}$, $j \leq 4$ in terms of $J_{m\varphi}$, $J_{m\rho}$, $J_{mz}$, which one could obtain by taking the first integral by $\widehat{\mathbf{p}}$ and the second only by $\varphi$ in \ref{eq:99} and \ref{eq:100}.
These expressions are written in cylindrical coordinates and should be substituted in COMSOL${ }^{\circledR}$ and then integrated by the nanoparticle's ``surface'' (integral by the rest spherical coordinates $\rho$ and $\theta_0$).

As a result, for example, for magnetic dipoles, one can
obtain:
\begin{equation}
    \begin{aligned}
        b_{1 -1} & =  \int_{\Omega} \dd\rho \dd \theta_0 \  j_1(k_0r) k_0^2 \frac{\sqrt{3 \pi }}{2} \rho \ \cdot \\ & \cdot
         ((-i J_{-1\varphi} + J_{-1\rho})\cos\theta_0-J_{-1z}\sin\theta_0) \\
        b_{1 0} & =  - i\int_{\Omega} \dd\rho \dd \theta_0 \  j_1(k_0r) J_{0\varphi} k_0^2 \sqrt{\frac{3 \pi}{2}}\rho \sin\theta_0\\
        b_{1 1} & =  \int_{\Omega} \dd\rho \dd \theta_0  j_1(k_0r) k_0^2 \frac{\sqrt{ 3 \pi }}{2} \rho \  \cdot \\ & \cdot
         ((i J_{1\varphi} + J_{1\rho})\cos\theta_0-J_{1z}\sin\theta_0),
    \end{aligned}
\end{equation}
where the integration should be performed over the nanoparticle's volume,  which appears as a surface in 2D geometry; $\theta_0$ is the polar (zenith) angle in the spherical coordinate system.

\begin{figure}[t]
\includegraphics[width=0.47\textwidth]{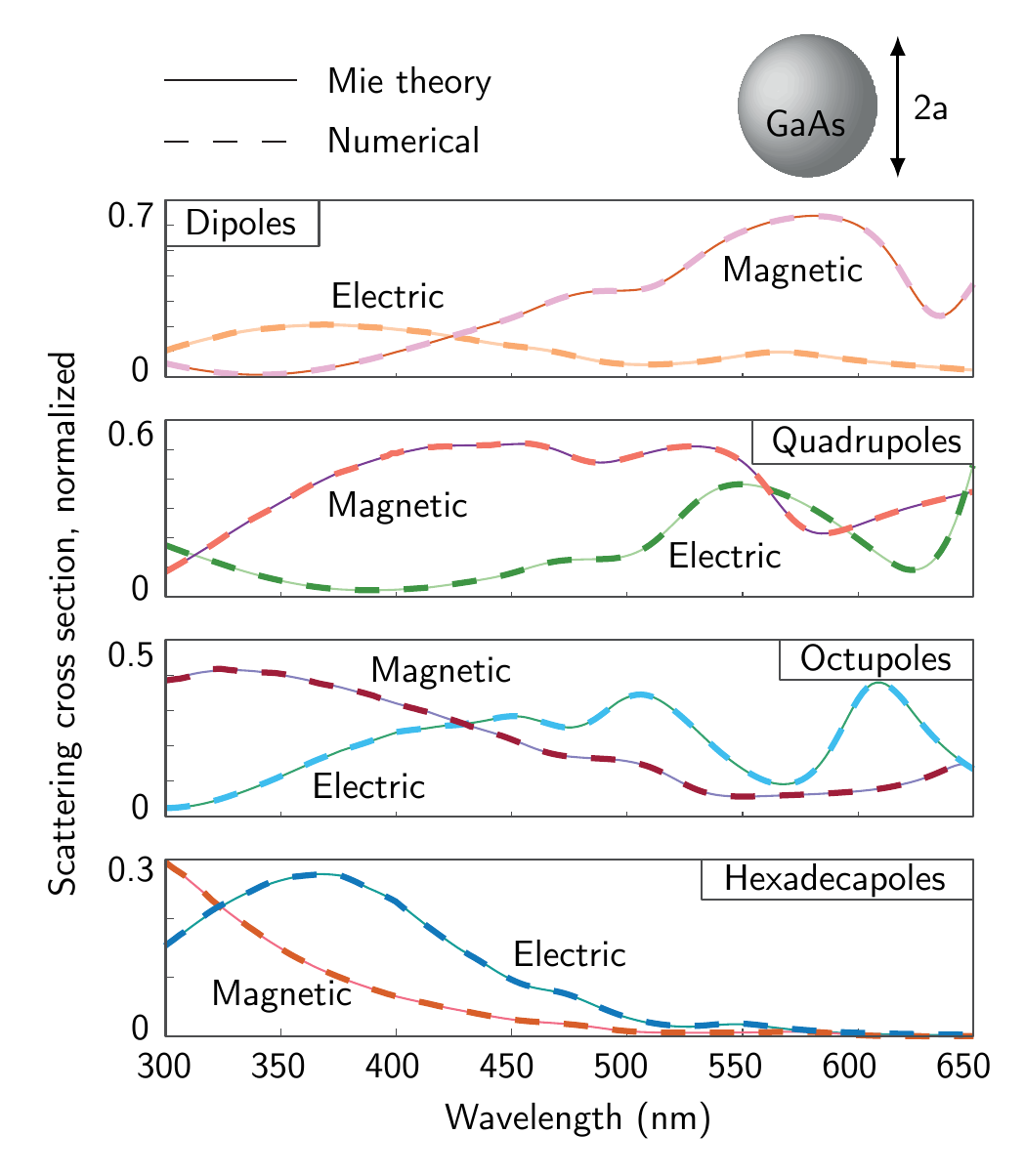}
\caption{Partial cross-sections of a plane wave scattered on a sphere corresponding to each multipole moment and normalized over the geometrical $\sigma_{\text{geom}}=\pi a^2$ (radius of sphere $a=250$~nm ) calculated with the exact expressions~\eqref{eq:99}--\eqref{eq:100} (sold lines) and with Mie theory (dashed lines). Spherical particle made for GaAs has radius of $a=250$ nm. }
\label{fig:multdecomp}
\end{figure}

We used the derived expressions of the multipolar moments for the case of light scattering on a sphere.
The comparison of the extracted multipoles via  Eqs.~(\ref{eq:99}--\ref{eq:100}) with the analytical results predicted by the Mie theory~\cite{BibEntry2023Jul} is shown in Fig.~\ref{fig:multdecomp} for the case of GaAs sphere of radius $a=250$ nm placed in a free space.
One can see excellent agreement between the Mie theory and numerical simulations with account for axial symmetry of the structure.
The  COMSOL Multiphysics$^{\mbox{\scriptsize\textregistered}}$ 
\href{https://github.com/Sag050196/multipole-decomposition-and-second-harmonic-generation-using-2D.git}{file} reproducing the results shown in Fig.~\ref{fig:multdecomp} is available~\cite{gitmultipole}.
Note that the total scattering cross-section can be obtained by summing over all multipolar contributions.

\section{Second harmonic generation}
\label{sec:SHG}
\begin{figure}[ht]
\begin{center}
    \includegraphics[width=0.42\textwidth]{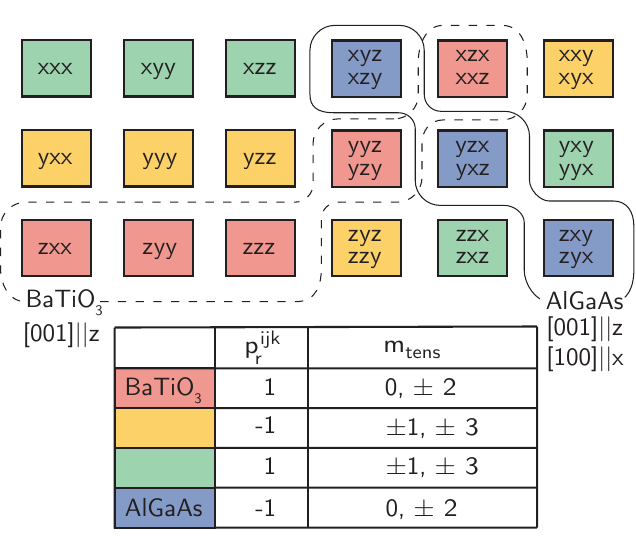}
    \caption{Possible $m_{\text{tens}}$ for different {$\hat{\chi}^{(2)}$} tensor components. Parity $p_r^{ijk}$ is also given, which reflects the behavior of the tensor under reflection in $y=0$ plane. This affects the second-harmonic parity under this reflection~\cite{Frizyuk2019-Second-harmonicgene} but does not play major role for our considerations. 
    {Note, that in our COMSOL\textsuperscript{\textregistered}   \href{https://github.com/Sag050196/multipole-decomposition-and-second-harmonic-generation-using-2D.git}{model file}~\cite{gitmultipole} these components are also marked by their colors for convenience.}} \label{figtens}
\end{center}
\end{figure}
In this section, we will extended method to speed up and improve the performance of simulations of second-harmonic generation (SHG) from subwavelength scatterers.
The second harmonic generation is a nonlinear optical process of interaction of two photons of the same frequency and  generation of the third photon with doubled frequency~\cite{Boyd2008-NonlinearOptics}.
From the early years of nonlinear optics, second and higher harmonic generation was rightly regarded as an effective tool for frequency conversion. 
Meanwhile, the developing efficient subwavelength sources of SHG is still one of the topical problems of experimental and theoretical nanophotonics \cite{Zalogina2023-High-harmonicgenerat, Koshelev2020-Subwavelengthdielect, Fedorov2020Aug, Saerens2020Sep, Bernasconi2016-Modeanalysisofseco, McLaughlin2022-Nonlinearopticsing, Cheng2021-SuperscatteringSupe, Camacho-Morales2016-NonlinearGeneration, deBeer2009-NonlinearMietheory, Liu2022-Third-andSecond-Har}.
Solution of the SHG problem even in the simplest geometries such as  spherical scatterer and plane wave excitation (Mie geometry) is a complex problem~\cite{Dadap2004Jul, Frizyuk2019-Second-harmonicgene, Butet2012Aug}, and numerical methods play a crucial role in designing nanophotonic systems.
The axial symmetry of the scatterers allows for significantly speeding up the simulations of the SHG by using azimuthal expansion method.

The main challenges of the extension of the proposed method to the second-harmonic domain are 
i) the nonlinearity of the problem and 
ii) the symmetry of the material tensor responsible for SHG.
Indeed, one can describe generation at the doubled frequency $2\omega$ via the polarization vector $\mathbf{P}^{2 \omega}$ determined by 
the second-order nonlinear optical susceptibility tensor $\hat \chi^{(2)}$:
\begin{equation}
\label{eq:sus_tensor_definition}
    \mathbf{P}^{2 \omega}(\mathbf{r}) = \varepsilon_0 \hat{\chi}^{(2)} \mathbf {E}^{\omega}(\vb{r}) \mathbf{E}^{\omega}(\vb{r}).
\end{equation}
where $\varepsilon_0$ is vacuum permittivity.
This approach is valid for non-centrosymmetric materials such as gallium arsenide~\cite{Rocco2020-VerticalSecondHarmo, Timofeeva2018-AnapolesinFree-Stan, Gigli2020-Quasinormal-ModeNon, Carletti2017-Controllingsecond-ha, Gigli2019-All-DielectricNanore, deCeglia2019-Second-HarmonicGener} or lithium niobate~\cite{Carletti2019-Secondharmonicgener, Ma2021-NonlinearLithiumNio, Park2022-High-efficiencysecon, Hao2020-Second-harmonicgener, Rao2018-Second-HarmonicGener, Ali2023-Near-fieldenhancemen, Fedotova2022-LithiumNiobateMeta}, and we will limit our consideration to them in this paper.

Already at this point, it becomes clear that despite of cylindrical symmetry of the scatterer, the nonlinearity of the problem and the symmetry of the tensor ``mixes'' the input harmonics with different $m$ \cite{Nikitina2023-Nonlinearcirculardi, Frizyuk2019-Second-harmonicgene}.
One should not be discouraged by this fact, 
{our method still} can be applied here once proper expansion of the nonlinear tensor is carried out and {the incident field's frequency is doubled.}

First of all, let us expand the field inside the nanoparticle as follows 
\begin{equation}
    \label{eq:exp_inc}
    \ve E^{\text{in}} (\rho, \varphi, z) = \sum_{m = -\infty}^{\infty} \ve E_m^{\omega} (\rho,z) e^{-i m \vphi}.
\end{equation}
Now one can move on to the second-order susceptibility tensor, which for convenience we represent as follows:
\begin{equation}\label{eq:22}
    \hat{\chi}^{(2)} = {\chi}^{(2)}_{ijk} \vec e_i \otimes \vec e_j \otimes \vec e_k,
\end{equation}
where $\vec e_{i,j,k}$ is the unit vector with $i$, $j$, $k$ being $x$, $y$ or $z$.
Hereinafter, we will omit the sign of the tensor product.
After one introduces three variables $m_1$, $m_2$, $m_3$, where $m_1$ and $m_2$ are associated with the incident field and $m_3$ corresponds to the SHG field, the expansion~\eqref{eq:exp_inc} substituted into~\eqref{eq:sus_tensor_definition} gives
\begin{equation}
\begin{aligned}
     \label{eq: pm3}
     \vec P^{2\omega}(\rho, \varphi, z) = &
     \sum_{m_3}\vec P^{2\omega}_{m_3}(\rho, z)e^{-im_3\varphi} =  \\ & = 
 \vec e_i  \varepsilon_0{\chi}^{(2)}_{ijk}(\vec e_j\vec E^{\text{in}}(\vec r)) ( \vec e_k \vec E^{\text{in}}(\vec r)) = \\ =
     \sum_{m_1,m_2}\vec e_i \varepsilon_0\chi_{ijk}^{(2)} & (\vec e_j\vec E_{m_2}^{\omega}(\rho, z)e^{-im_2\varphi}) \cdot \\ 
     \cdot & ( \vec e_k \vec E_{m_1}^{\omega}(\rho, z)e^{-im_1\varphi}),
\end{aligned}
\end{equation}
where we use Einstein summation convention for $i, \ j, \ k$.

Importantly, here $m_3\ne m_1+m_2$ in the general case because of the spatial symmetry of  $\hat\chi^{(2)}$. 
Indeed, the relations for the cylindrical coordinate system
\begin{equation}\begin{cases}
    \vec{e}_{x} = \vec{e}_{\rho}\cos\varphi - \vec{e}_{\varphi}\sin\varphi, \\
    \vec{e}_{y} = \vec{e}_{\rho}\sin\varphi + \vec{e}_{\varphi}\cos\varphi, \\
    \vec{e}_{z} = \vec{e}_{z}
\end{cases}\end{equation} substituted into tensor components \eqref{eq:22} lead to the additional exponential terms.
For example, for $\vec{e}_{x}\vec{e}_{y}\vec{e}_{z}$  and $\vec{e}_{y}\vec{e}_{x}\vec{e}_{z}$ (we consider these two terms simultaneously for further convenience), one can obtain: 
\begin{equation}
\begin{aligned}
    \vec{e}_{x}\vec{e}_{y}\vec{e}_{z} =
    \frac{e^{2i\varphi} - e^{-2i\varphi}}{4i}\vec{e}_{\rho}\vec{e}_{\rho}\vec{e}_{z}-
    \frac{e^{2i\varphi} - e^{-2i\varphi}}{4i}\vec{e}_{\varphi}\vec{e}_{\varphi}\vec{e}_{z} + \\ + 
    \frac{e^{2i\varphi} + e^{-2i\varphi}-2}{4}\vec{e}_{\varphi}\vec{e}_{\rho}\vec{e}_{z} +
    \frac{e^{2i\varphi} + e^{-2i\varphi}+2}{4}\vec{e}_{\rho}\vec{e}_{\varphi}\vec{e}_{z}, \label{xyz}
\end{aligned}
\end{equation}
\begin{equation}
\begin{aligned}
    \vec{e}_{y}\vec{e}_{x}\vec{e}_{z} =
    \frac{e^{2i\varphi} - e^{-2i\varphi}}{4i}\vec{e}_{\rho}\vec{e}_{\rho}\vec{e}_{z} -
    \frac{e^{2i\varphi} - e^{-2i\varphi}}{4i}\vec{e}_{\varphi}\vec{e}_{\varphi}\vec{e}_{z} +\\+
    \frac{e^{2i\varphi} + e^{-2i\varphi}+2}{4}\vec{e}_{\varphi}\vec{e}_{\rho}\vec{e}_{z}+
    \frac{e^{2i\varphi} + e^{-2i\varphi}-2}{4}\vec{e}_{\rho}\vec{e}_{\varphi}\vec{e}_{z}.\label{yxz}
\end{aligned}
\end{equation}
Note that these expressions are still purely real, but we use the complex form to emphasize how the momentum projection changed due to the lattice symmetry.
One can see that the tensor components contain exponential factors, that we also need to take into account in Eq.~\eqref{eq: pm3}.
For them we will use the notation $e^{ im_{\text{tens}}\varphi}$. 
Therefore, angular momentum conservation does not work in a usual way as for cylindrical symmetry.
We would like to emphasize that the orientation of the crystal lattice is taken into account automatically, since it affects only the values of the  $\hat\chi^{(2)}$ tensor components~\cite{Nikitina2023-Nonlinearcirculardi}.
Note that the consideration should be different for materials with central symmetry~\cite{Smirnova2016-Multipolarnonlinear}; however, we believe that our approach is expandable to the latter case as well.

The  $m_{\text{tens}}$ for different components is given in Fig.~\ref{figtens}.
One can derive this by considering  the behavior of the unit vectors under rotations around the $z$-axis.
$\vec e_z$ is not transformed, so it does not contribute, and $\vec e_x$ and $\vec e_y$ provide $m=\pm 1$.
The parity under reflection in $y=0$ plane $p_r^{ijk}$ is also given, but this can only affect the selection rules~\cite{Frizyuk2019-Second-harmonicgener} and thus does not play major role in our considerations now. 
Therefore, Eq.~\eqref{eq: pm3} provides all possible non-zero nonlinear $\vec P_{m_3}^{2\omega}(\rho, z)e^{-im_3\varphi}$, where $m_3$ satisfies the following condition   
\begin{equation}
    \label{eq: cond}
    m_1+m_2+m_{\text{tens}}=m_3,
\end{equation}
and also takes into account the value of $m_{\text{tens}}$ for different tensor components. 

Let us consider this approach using the example of the [100] orientation GaAs tensor, which has only off-diagonal components ($\chi_{i j k}^{(2)}$ vanishes if any of the two indices $i, j, k$ coincide, and all other components are equal to each other)~\cite{Boyd2008-NonlinearOptics}.
We rewrite the tensor in cylindrical coordinate system according to Eq.~\eqref{xyz} and  substitute it into Eq.~\eqref{eq: pm3}.
It turns out that the additional momentum projection $m_{\text{tens}}=0$ vanishes after the summation of the $xyz+yxz$, $zxy+zyx$, $xzy+yzx$ components.
After that, we can write all possible components $\rho, \varphi$ and $z$ of the induced polarization as follows. As an example, we provide the $xyz+yxz$-term below, while $zxy+zyx$ and $xzy+yzx$ terms can be obtained similarly:
\begin{widetext}
\begin{equation}
\begin{aligned}
    P^{2\omega}_{\rho, m_3}e^{-im_3\varphi} =   \varepsilon_0(\chi^{(2)}_{yxz} + \chi^{(2)}_{xyz})\sum_{m_1, m_2, m'_1, m'_2}2\left[e^{-i(m_1+m_2+2)\varphi}\left(-\frac{1}{2i}E_{\rho,m_2}^{\omega} E_{z,m_1}^{\omega}+\frac{1}{2}E_{\varphi,m_2}^{\omega} E_{z,m_1}^{\omega}\right) + \right.\\ +
    \left. e^{-i(m'_1+m'_2-2)\varphi}\left(\frac{1}{2i}E_{\rho,m'_2}^{\omega} E_{z,m'_1}^{\omega} + \frac{1}{2}E_{\varphi,m'_2}^{\omega} E_{z,m'_1}^{\omega}\right)\right], 
\end{aligned} 
\end{equation}
\begin{equation}
\begin{aligned}
    P^{2\omega}_{\varphi, m_3}e^{-im_3\varphi} =   \varepsilon_0(\chi^{(2)}_{yxz} + \chi^{(2)}_{xyz})\sum_{m_1, m_2, m'_1, m'_2}2\left[e^{-i(m_1+m_2+2)\varphi}\left(\frac{1}{2i}E_{\varphi,m_2}^{\omega} E_{z,m_1}^{\omega}+\frac{1}{2}E_{\rho,m_2}^{\omega} E_{z,m_1}^{\omega}\right)+\right. \\+ 
    \left. e^{-i(m'_1+m'_2-2)\varphi}\left(-\frac{1}{2i}E_{\varphi,m'_2}^{\omega} E_{z,m'_1}^{\omega}+\frac{1}{2}E_{\rho,m'_2}^{\omega} E_{z,m'_1}^{\omega}\right)\right], 
\end{aligned} 
\end{equation}
\begin{equation}
\begin{aligned}
    P^{2\omega}_{z, m_3}e^{-im_3\varphi} =   \varepsilon_0(\chi^{(2)}_{yxz}+\chi^{(2)}_{xyz})\sum_{m_1, m_2, m'_1, m'_2}\left[e^{-i(m_1+m_2+2)\varphi}\left(-\frac{1}{2i}E_{\rho,m_2}^{\omega} E_{\rho,m_1}^{\omega}+\frac{1}{2i}E_{\varphi,m_2}^{\omega} E_{\varphi,m_1}^{\omega}\right.+\right. \\+ 
    \left. \left. \frac{1}{2}E_{\varphi,m_2}^{\omega} E_{\rho,m_1}^{\omega}+\frac{1}{2}E_{\rho,m_2}^{\omega} E_{\varphi,m_1}^{\omega}\right)\right. + \left.
    e^{-i(m'_1+m'_2-2)\varphi}\left(\frac{1}{2i}E_{\rho,m'_2}^{\omega} E_{\rho,m'_1}^{\omega}-\frac{1}{2i}E_{\varphi,m'_2}^{\omega} E_{\varphi,m'_1}^{\omega}\right.+\right. \\+ 
    \left. \left. \frac{1}{2}E_{\varphi,m'_2}^{\omega} E_{\rho,m'_1}^{\omega}+\frac{1}{2}E_{\rho,m'_2}^{\omega} E_{\varphi,m'_1}^{\omega}\right) \right]. 
\end{aligned} 
\end{equation}
\end{widetext}
Note that even if each separate $\hat \chi{^{(2)}}$ component contains $m_{\text{tens}}=0$ [see~Eqs.~\eqref{xyz} and~\eqref{yxz}], it disappears after summation by pairs $xyz+yxz$ and so on.
This is specific for GaAs and will generally not happen.
\begin{figure}[t]
    \includegraphics[width=0.47\textwidth]{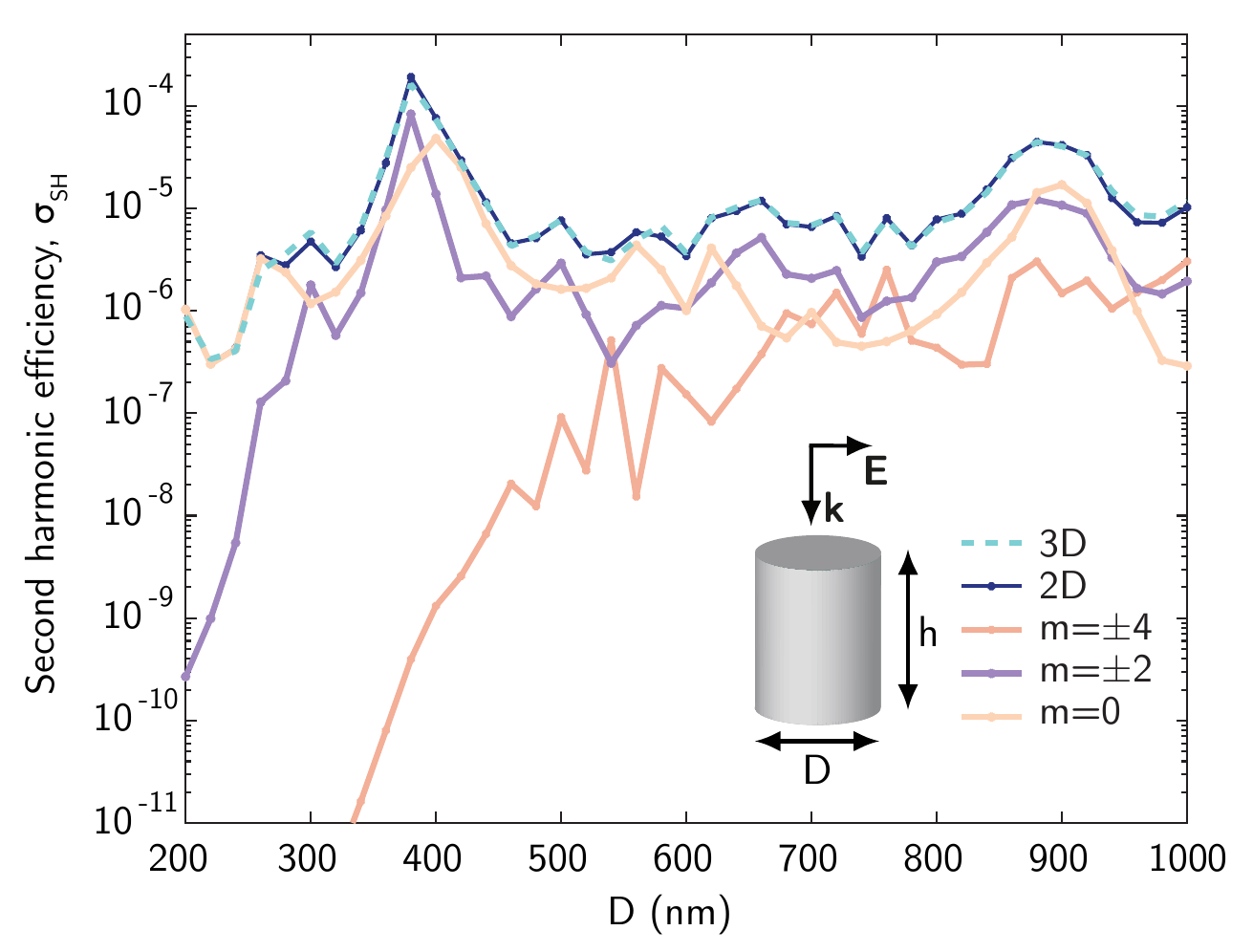}
    \caption{Dependence of the second-harmonic efficiency
    in GaAs cylinder of height $h=400$~nm excited by a normally incident plane wave at $\lambda = 1550$~nm  on the cylinder diameter calculated by the 3D model and 2D model. Contributions calculated in the 2D model for different orders $m\in \{0,\pm2,\pm4\}$ are shown. 
    } 
\label{fig:SHG}
\end{figure}
\begin{figure*}[t!]
    \includegraphics[width=0.99\textwidth]{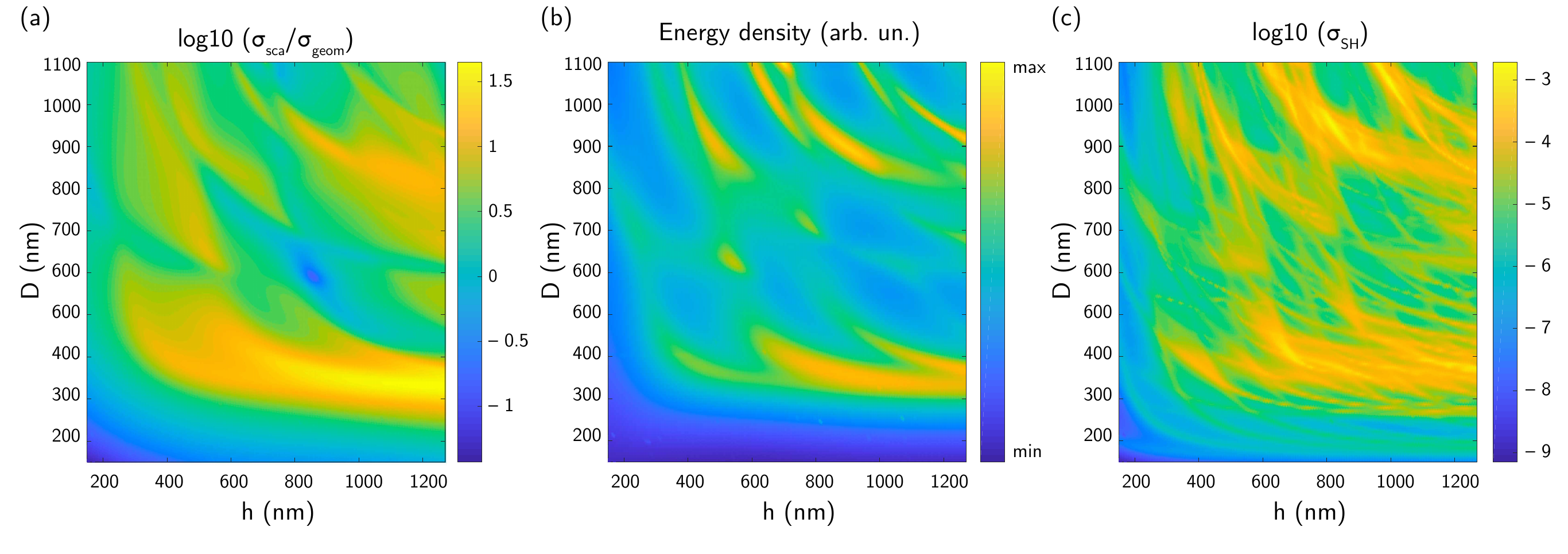}
    \caption{Base-10 logarithmic scale maps for (a) normalised scattering cross-section  $\sigma_{\text{sca}}$, (b) average electro-magnetic energy inside the nanostructure $\sigma_{\text{wav}}$, integrated over the nanoparticle volume, and normalized by the volume as well as (c) SHG $\sigma_{\text{SHG}}$ for the GaAS nanocylinder as a function of its height $h$ and diameter $D$, obtained in the two-dimensional model. Wavelength of the incident wave $\lambda=1550$~nm.}
    \label{fig:SHGmap}
\end{figure*}

During the simulations, the range of $m_3$ should be chosen manually, based on the selection rules~\cite{Frizyuk2019-Second-harmonicgene} and requires attention.
Note that despite possible values are 
$|m_3| \leq 2M_{\text{max}}+|m_{\text{tens}}|$,
we recommend to choose the maximum value 
$|m_3| \leq M_{\text{max}}$
to preserve the accuracy, and check if there are resonances with such $m$ in this range. 
Since one can take only the finite number of harmonics in the numerical simulation, we assume that the numbers $m_{1,2}$ are in the range $m_{1,2}\in [-M_{\text{max}}.. M_{\text{max}}]$. Thus, one should take all possible values from this range, which together with $m_{\text{tens}}$  satisfy~\eqref{eq: cond}.
So one should  impose the following conditions: 
 \begin{enumerate}
     \item $m_2=m_3-m_1-m_{\text{tens}} $
     \item {$m_1 \in [\max(m_3-M_{\max}-m_{\text{tens}}, -M_{\max})\dots\\ \dots \min(m_3+M_{\max}-m_{\text{tens}}, M_{\max})]$.}
 \end{enumerate}
We used the described method to simulate the SHG in the GaAs nanocylinder oriented such as $[100]||x$, $[001]||z$.
According to the selection rules for a normally incident linearly polarized wave, the nonlinear response corresponding to the second harmonic generation will be nonzero for 
$m_3 \in \{0,\pm2,\pm4\}$~\cite{Frizyuk2019-Second-harmonicgener}.
This happens because $m=\pm 1$ for the incident field, which contributes twice and leads to $m\in \{0, \pm 2\}$, and $m_{\text{tens}}=\pm 2$.
Figure~\ref{fig:SHG} shows the dependence of the nonlinear signal intensity on the cylinder diameter while the its height is fixed at $h=400$ nm.
The figure depicts the SHG cross-section $\sigma_\text{SH}$, which is defined as the second harmonic intensity normalized over the geometric cross-section and the intensity of the incident radiation~\cite{Frizyuk2019-Second-harmonicgener}.
The excitation plane wave is incident along the cylinder axis  at the wavelength $\lambda = 1550$~nm.
We have compared the results of the full 3D simulations and simulation with the proposed  2D axisymmetric problem solution, which shows an excellent agreement proving the correctness of our method.  
In addition, the contributions of different polarization components with order $m\in \{0,\pm2,\pm4\}$ are demonstrated. 

The high performance and computational efficiency of the method allows for sweeping over the large sets of parameters.
As an example, Figure~\ref{fig:SHGmap}(a) shows maps of linear scattering cross-section for various cylinder heights and diameters. 
Along with the scattering cross-section, the average  electromagnetic energy density inside the cylinder and  SHG cross-section are also shown in Figs.~\ref{fig:SHGmap}(b) and \ref{fig:SHGmap}(c).
For the convenience of the readers, the  COMSOL Multiphysics\textsuperscript{\textregistered}   \href{https://github.com/Sag050196/multipole-decomposition-and-second-harmonic-generation-using-2D.git}{model file}~\cite{gitmultipole} is available for downloading.
The model allows for obtaining the results shown in Figs.~\ref{fig:SHG} and~\ref{fig:SHGmap}.

\section{Discussion} \label{sec:analasis}

\begin{figure}[t]
\includegraphics[width=0.47\textwidth]{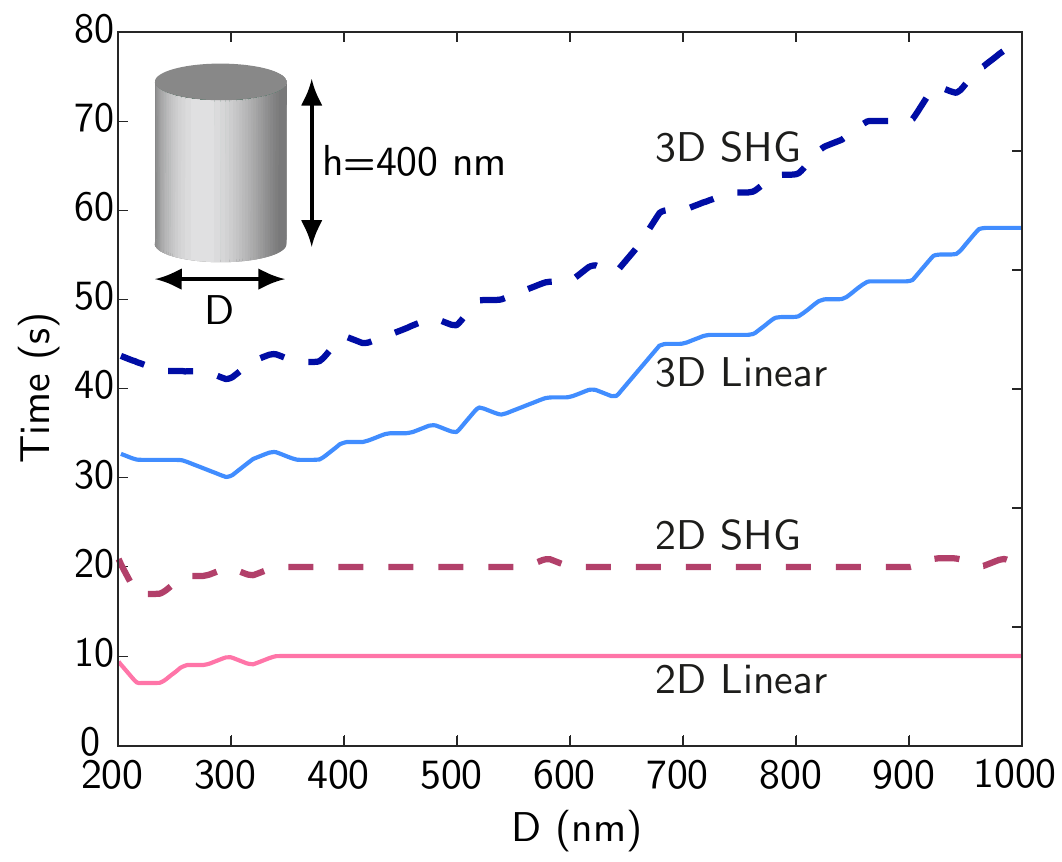}
\caption{Linear scattering and SHG computation time for 2D  an 3D models depending on the radius of GaAs cylinder with height of 400 nm upon excitation of the plane wave at $\lambda = 1550$ nm.}
\label{fig:time}
\end{figure}
\begin{figure*}[t]
\includegraphics[width=0.99\textwidth]{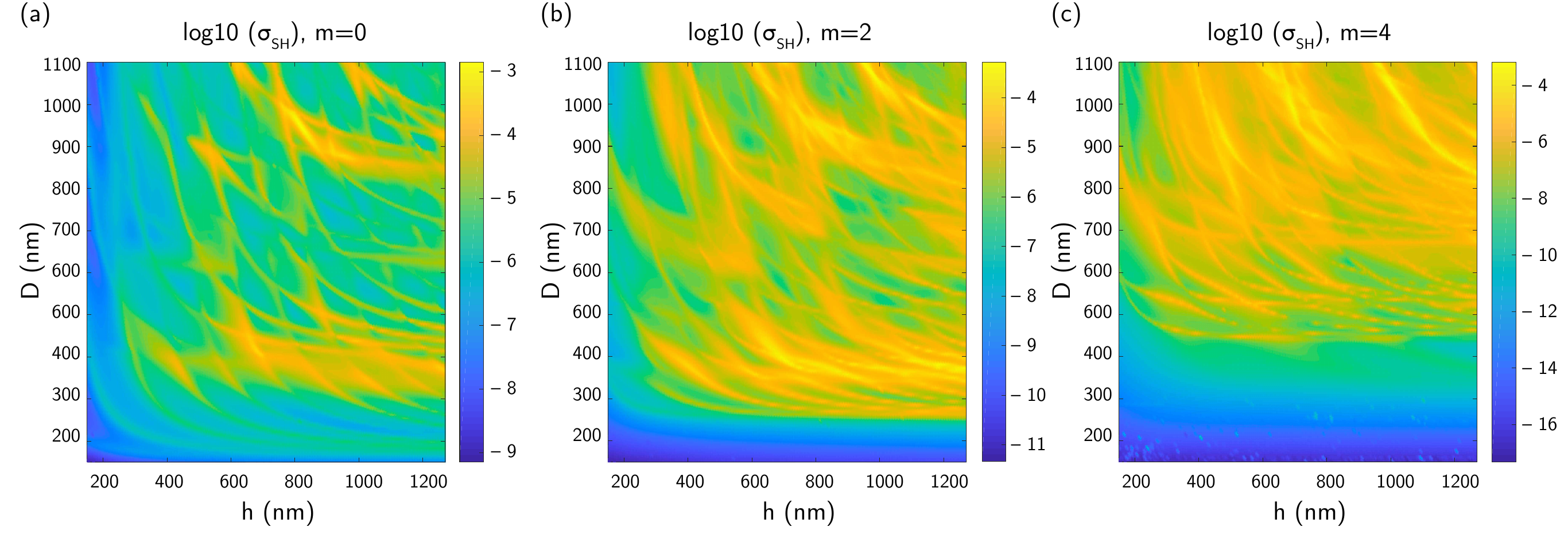}
\caption{Base-10 logarithmic scale maps for different $m$ contibution ((a) $m=0$, (b) $m=2$, (c) $m=4$) into the SHG $\sigma_{\text{SHG}}$ for the GaAS nanocylinder, as a function of its height $h$ and diameter $D$, obtained in the two-dimensional model. Wavelength of the incident wave $\lambda=1550$~nm. }
\label{fig:SHGmap2}
\end{figure*}

Finally, let us discuss the advantage in computational time and required resources
that the proposed method provides.
It allows to reduce 3D problem to a set of 2D problems which could be simulated much faster.
Though, one need to perform a number of 2D simulations  proportional to the number of required harmonics, 
$2M_{\text{max}}+1$ in linear case and number of a chosen $m_3$ in SHG case, 
it still appears to be much faster if the computation area and/or the number of mesh elements in 3D model are large. 
This benefit is illustrated by Fig.~\ref{fig:time}, where we compare the elapsed time to simulate scattering on the dielectric cylinder for full 3D and 2D axisymmetric geometry.
The simulation time is shown for both linear scattering and SHG simulations.
Note that the SHG computation time is longer because it includes the solution of the linear problem.
One can see that the computation time of the 2D model is much lower than for the 3D model and almost does not change with the size of the modeled object (cylinder diameter $D$), while  time required for 3D simulation rapidly increases. 

Another advantage is that the our method immediately provides extra information on the particular Fourier harmonic contribution.
This is often helpful for further analysis of the simulation results such as multipole decomposition or far-field radiation patterns.
Indeed, Fig.~\ref{fig:SHGmap2} shows the decomposition of the SH map shown in Fig.~\ref{fig:SHGmap} over the Fourier harmonics with different numbers $m$. 
One can see that the such a decomposition immediately explains origins of the peaks in the SHG intensity spectra.

Finally, the  proposed method of SHG simulations can be extended to other nonlinear processes such as third  harmonic generation.
While the general approach will be exactly the same, the main difference will be in the expansion of the third-order nonlinear tensor into the Fourier series over $e^{-im\varphi}$ and accounting for three input fields in nonlinear polarization tensor [see Eq.~\eqref{eq:sus_tensor_definition}]. 

\section{Conclusions}
\label{sec:conclusions}

In conclusion, this work proposes novel efficient numerical tool based on COMSOL Mutliphysics\textsuperscript{\textregistered} software for  simulating linear and nonlinear light scattering from the nanophotonic scatterers of cylindrical symmetry.
Taking the advantage of the symmetry of the problem, one can reduce simulations from 3D problem to a set of 2D problems, which can be computed  much faster. 
We provide the particular expressions for performing   multipolar decomposition of the scattered fields and computing optical forces acting on the scatterers.
We also showed that the proposed method is efficient for simulating second harmonic generation.
We showed that it gives sufficient benefit in computational time when simulating the second-harmonic generation from resonant dielectric nanocylinders made of GaAs. 

We also provided detailed description of the method and accompanied it with the COMSOL Multiphysics\textsuperscript{\textregistered} sample models freely available for downloading.
We believe that the proposed  numerical tool represents a significant advancement in the simulating linear and nonlinear scattering from axially symmetric structures.
Its computational efficiency, accuracy and versatility make it a valuable asset for researchers in this field, allowing a deeper understanding and facilitating the design of novel nanophotonic devices.
\section*{Acknowledgements} \label{sec:acknowledgements}
We thank Kirill Koshelev for fruitful and valuable discussions. The work was supported by the Russian Science Foundation (22-12-00204). A.B. and M.P. acknowledge the Federal Academic Leadership Program Priority 2030.  

\normalem
\bibliography{Artikel}

\appendix*

\end{document}


\title{Supplemental Material: \\ 
Use of symmetry or how to make the scattering calculations in COMSOL faster}

\author{Sergei Gladyshev}
\altaffiliation{School of Physics and Engineering, ITMO University, 191002 St. Petersburg, Russia}

\author{Olesia Pashina}
\altaffiliation{School of Physics and Engineering, ITMO University, 191002 St. Petersburg, Russia}

\author{Alexey Proskurin}
\altaffiliation{School of Physics and Engineering, ITMO University, 191002 St. Petersburg, Russia}

\author{Anna Nikolaeva}

\author{Zarina Sadrieva}
\altaffiliation{School of Physics and Engineering, ITMO University, 191002 St. Petersburg, Russia}

\author{Andrey Bogdanov }
\altaffiliation{School of Physics and Engineering, ITMO University, 191002 St. Petersburg, Russia}
\altaffiliation{Qingdao Innovation and Development Center of Harbin Engineering University, 266404 Qingdao, China}
\email{a.bogdanov@metalab.ifmo.ru}

\author{Mihail Petrov}
\altaffiliation{School of Physics and Engineering, ITMO University, 191002 St. Petersburg, Russia}
\email{m.petrov@metalab.ifmo.ru}

\author{Kristina Frizyuk}
\altaffiliation{School of Physics and Engineering, ITMO University, 191002 St. Petersburg, Russia}
\email{k.frizyuk@metalab.ifmo.ru}

\begin{abstract}
    Section I shows how the incident TM-polarized plane wave is expanded into a series of cylindrical waves. Section II shows how the Cartesian components of optical forces can be found using the Maxwell stress written in cylindrical coordinates. Section III shows how the absorption, extinction, and scattering cross-sections can be calculated in cylindrical coordinates via surface and volume integration. Section IV contains the components of the second-order nonlinear susceptibility tensor in cylindrical coordinates.    
\end{abstract}

\maketitle



\section{S1. Cylindrical wave expansion of TM-polarized plane wave}


 Let's consider the incidence of a plane wave on a cylindrical object at an angle of incidence $\theta$ with TM-polarization. We have that  the $\vb{k}_0$-vector lies in the xz-plane, while $\vb{H}$-field has only $y$-component.

The incident electrical field in cylindrical coordinates will have the form 
\begin{equation}
\vb{E}^{\text{inc}}=\left(\begin{array}{c}
E_{\rho}^{\text{inc}} \\
E_{\varphi}^{\text{inc}} \\
E_{z}^{\text{inc}}
\end{array}\right)=\left(\begin{array}{c}
E_{0} \sin \theta \cos \varphi \\
E_{0} \sin \theta \sin \varphi \\
E_{0} \cos \theta
\end{array}\right) e^{i k_{0 z} z-i k_{0 x} \rho \cos \varphi}
\end{equation}

The radial component can be expressed as follows:

\begin{equation}\label{eq:rho_tm}
E^{\text{inc}}_\rho = \sum_{m=-\infty}^{+\infty} E_{0} \sin \theta e^{i k_{0 z} z}(-i)^{m-1}  \frac{\dd J_{m}(k_{0x}\rho)}{\dd (k_{0x}\rho)} e^{-i m \varphi}
\end{equation}

The azimuthal component has the form:

\begin{equation}\label{eq:varphi_tm}
E^{\text{inc}}_\varphi = \sum_{m=-\infty}^{+\infty} E_{0} \sin \theta e^{i k_{0 z} z}(-i)^{m+2} \frac{m}{k_{0x} \rho} J_{m}\left(k_{0 x} \rho\right) e^{-i m \varphi}
\end{equation}

The $z$-component has the form:

\begin{equation}\label{eq:z_tm}
E^{\text{inc}}_z = \sum_{m=-\infty}^{+\infty} E_{0} \cos \theta e^{i k_{0 z} z}(-i)^{}  J_{m}\left(k_{0 x} \rho\right) e^{-i m \varphi}
\end{equation}

In particular, the $\rho$-component and $z$-component are even:

\begin{equation}
\begin{aligned}
    E_{m, \rho}= E_{0} \sin \theta e^{i k_{0 z} z}(-i)^{m-1}  \frac{\dd J_{m}(k_{0x}\rho)}{\dd (k_{0x}\rho)} = E_{-m, \rho}
\end{aligned}    
\end{equation}

\begin{equation}
\begin{aligned}
    E_{m, z}=  E_{0} \cos \theta e^{i k_{0 z} z}(-i)^{}  J_{m}\left(k_{0 x} \rho\right) = E_{-m, z}
\end{aligned}    
\end{equation}

while the $\varphi$-component is odd:

\begin{equation}
\begin{aligned}
    E_{m, \varphi}= E_{0} \sin \theta e^{i k_{0 z} z}(-i)^{m+2} \frac{m}{k_{0x} \rho} J_{m}\left(k_{0 x} \rho\right) = E_{-m, \varphi}
\end{aligned}    
\end{equation}



\section{S2. Components of the Maxwell stress tensor}

\begin{figure}[h]
\includegraphics[width=0.47\textwidth]{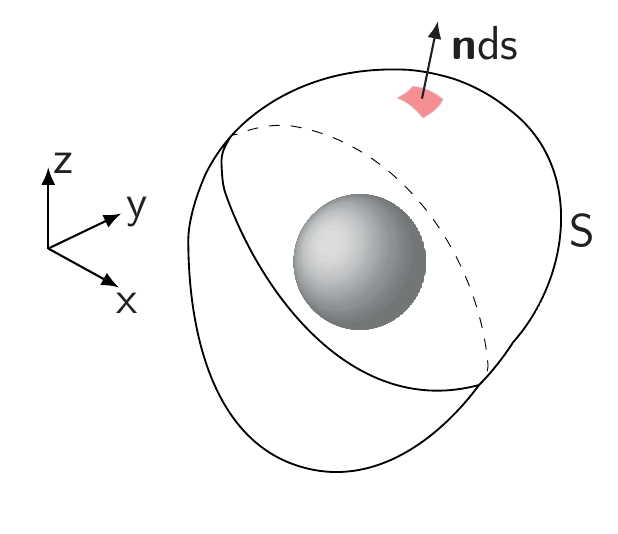}
\caption{Calculation of the optical force via integration of the Maxwell stress tensor.}
\label{fig:supp3}
\end{figure}

Generally, one can find the optical force acting on a particle by integrating the Maxwell stress tensor over the surface enclosing the particle (see Fig.~\ref{fig:supp3}):
\begin{equation}\label{eq:maxwell_T}
    \mathbf{F} = \iint_{S}^{}{\widehat{T}\mathbf{n}}\, \mathrm{d}s,
\end{equation}
where the Maxwell stress tensor $\widehat T$ is given by
\begin{equation*}
    \widehat{T} = \mathbf{E} \otimes \mathbf{D} + \mathbf{B} \otimes \mathbf{H} - \frac{1}{2}\left( \mathbf{E}\mathbf{D} + \mathbf{B}\mathbf{H} \right)\widehat{I}.
 \end{equation*}
Here $\mathbf n$ is a normal vector to the surface $S$,  $\mathbf E$ and $\mathbf D$ are electric field and electric displacement field, respectively; $\mathbf B$ and $\mathbf H$ are magnetic fields; and $\widehat I$ is the identity tensor.

If the particle is in the vacuum (\(\varepsilon_{r} = 1\), \(\mu_{r} = 1\)),
\begin{equation}
    \widehat{T} = \varepsilon_{0}\mathbf{E} \otimes \mathbf{E} + \mu_{0}\mathbf{H} \otimes \mathbf{H} - \frac{1}{2}\left( \varepsilon_{0}\mathbf{E}\mathbf{E} + \mu_{0}\mathbf{H}\mathbf{H} \right)\widehat{I}
\end{equation}

This expression can be divided into two parts so that each part involves either electric or magnetic fields only. Hereinafter, we will focus on the electric part
\begin{equation}\label{eq:TE_def}
    \widehat T^E=\varepsilon_{0}\mathbf{E} \otimes \mathbf{E} - \frac{1}{2}  \varepsilon_{0}\mathbf{E}\mathbf{E} \widehat{I}
\end{equation}
only, while all the expressions for the magnetic part can be achieved from the electric one by the substitutions $\mathbf E\to\mathbf H$ and $\varepsilon_0\to\mu_0$. The resulting terms must be added up to get the total optical force acting on the particle.

When a considered system possesses rotational symmetry, it is more convenient and computationally efficient to solve Maxwell equations in cylindrical \((\rho,\varphi,z)\) coordinates. By contrast, the components of the optical force are commonly found in Cartesian $(x,y,z)$ coordinates. Therefore, there is a need for introducing the mixed ''Cartesian-cylindrical'' coordinates for the Maxwell stress tensor to ensure it allows to meet both requirements.

To describe the relationship between vector coordinates in Cartesian and cylindrical coordinate systems, one can use the transformation matrix $\widehat P$:
\begin{equation}
    \begin{pmatrix}
        E_{x} \\
        E_{y} \\
        E_{z} 
    \end{pmatrix} = 
    \underbrace{\begin{pmatrix}
        \cos\varphi & - \sin\varphi & 0 \\
        \sin\varphi & \cos\varphi & 0 \\
        0 & 0 & 1 
    \end{pmatrix}}_{\widehat P}
    \begin{pmatrix}
        E_{\rho} \\
        E_{\varphi} \\
        E_{z}
    \end{pmatrix} \equiv {\widehat{P}}_{\left( \mathbf{e_{x}},\mathbf{e_{y}},\mathbf{e_{z}} \right) \leftarrow \left( \mathbf{e_{\rho}},\mathbf{e_{\varphi}},\mathbf{e_{z}} \right)}\begin{pmatrix}
    E_{\rho} \\
    E_{\varphi} \\
    E_{z} \\
    \end{pmatrix}
\end{equation}

Let us rewrite~\eqref{eq:TE_def} by using the transformation matrix: 
\begin{multline}
    \widehat{T}^{E} = \varepsilon_0\left( \left( E_{x}\mathbf{e_{x}} + E_{y}\mathbf{e_{y}} + E_{z}\mathbf{e_{z}} \right) \otimes \left( E_{\rho}\mathbf{e_{\rho}} + E_{\varphi}\mathbf{e_{\varphi}} + E_{z}\mathbf{e_{z}} \right)-\frac{1}{2}\left( E_{\rho}^{2} + E_{\varphi}^{2} + E_{z}^{2} \right){\widehat{P}}_{\left( \mathbf{e_{x}},\mathbf{e_{y}},\mathbf{e_{z}} \right) \leftarrow \left( \mathbf{e_{\rho}},\mathbf{e_{\varphi}},\mathbf{e_{z}} \right)} \right).
\end{multline}
After applying the explicit formulas for $\mathbf{e}_i$ and performing some straightforward simplifications, one gets
\begin{equation}\label{eq:TE_rho}
\begin{pmatrix}
{{T^{E}}}_{x,\rho}\\
{{T^{E}}}_{y,\rho}\\
{{T^{E}}}_{z,\rho}
\end{pmatrix}
     = \varepsilon_0\begin{pmatrix}
        \frac{E_{\rho}^{2} - E_{\varphi}^{2} - E_{z}^{2}}{2}\cos\varphi - E_{\rho}E_{\varphi}\sin\varphi \\
        \frac{E_{\rho}^{2} - E_{\varphi}^{2} - E_{z}^{2}}{2}\sin\varphi + E_{\rho}E_{\varphi}\cos\varphi \\
        E_{\rho}E_{z}
    \end{pmatrix},
\end{equation}
\begin{equation}\label{eq:TE_phi}
    \begin{pmatrix}
{{T^{E}}}_{x,\varphi}\\
{{T^{E}}}_{y,\varphi}\\
{{T^{E}}}_{z,\varphi}
\end{pmatrix}
    = \varepsilon_0\begin{pmatrix}
        E_{\rho}E_{\varphi}\cos\varphi - \frac{E_{\varphi}^{2} - E_{\rho}^{2} - E_{z}^{2}}{2}\sin\varphi \\
        E_{\rho}E_{\varphi}\sin\varphi + \frac{E_{\varphi}^{2} - E_{\rho}^{2} - E_{z}^{2}}{2}\cos\varphi \\
        E_{\varphi}E_{z}
    \end{pmatrix},
\end{equation}
and
\begin{equation}\label{eq:TE_z}
    \begin{pmatrix}
{{T^{E}}}_{x,z}\\
{{T^{E}}}_{y,z}\\
{{T^{E}}}_{z,z}
\end{pmatrix}
= \varepsilon_0\begin{pmatrix}
        \left( E_{\rho}\cos\varphi - E_{\varphi}\sin\varphi \right)E_{z} \\
        \left( E_{\rho}\sin\varphi + E_{\varphi}\cos\varphi \right)E_{z} \\
        \frac{E_{z}^{2} - E_{\rho}^{2} - E_{\varphi}^{2}}{2},
    \end{pmatrix}
\end{equation}.

To account for the averaging over time, all the products $E_a E_b$ should be replaced with $\frac 12 \mathop{\mathrm{Re}}E_a E_b^*$.

Let us assume that the spatial distribution of the electric field is already calculated and given in the form
\begin{equation}
    E_a = \sum_{m = - \infty}^{\infty}{E_{m,a}e^{- im\varphi}},
\end{equation}
where $E_{m,a}=E_{m,a}(\rho,z)$ does not depend on the azimuthal angle $\varphi$. Then one can integrate~\eqref{eq:maxwell_T} over $\varphi$ since the orthogonality relations:
\begin{equation}
    \int_{0}^{2\pi}{E_{a}E_{b}^{*}\cos\varphi}\, \mathrm{d}\varphi = 2\pi \cdot \frac{1}{2}\sum_{m = - \infty}^{\infty}\left( E_{m + 1,a}\left( E_{m,b} \right)^{*} + E_{m,a}\left( E_{m+1,b} \right)^{*} \right)
\end{equation}
\begin{equation}
    \int_{0}^{2\pi}{E_{a}E_{b}^{*}\sin\varphi}\, \mathrm{d}\varphi = - 2\pi \cdot \frac{i}{2}\sum_{m = - \infty}^{\infty}\left( E_{m+1,a}\left( E_{m,b} \right)^{*} - E_{m,a}\left( E_{m+1,b} \right)^{*} \right)
\end{equation}
\begin{equation}
    \int_{0}^{2\pi}{E_{a}E_{b}^{*}}\, \mathrm{d}\varphi = 2\pi \cdot \sum_{m = - \infty}^{\infty}{E_{m,a}\left( E_{m,b} \right)^{*}}
\end{equation}

Therefore, one can get rid of 3D~integration in the evaluation of optical forces in an axially symmetrical system:
\begin{multline}
    \left\langle{F}_{x}\right\rangle = \frac{\varepsilon_0}{4}\mathop{\mathrm{Re}}\int_{C_R}^{}{2\pi\rho\, \mathrm{d}c}\,\sum_{m = - \infty}^{+ \infty}\left\lbrack \left( E_{m,\rho}\left( E_{m + 1,\rho} \right)^{*} - E_{m,\varphi}\left( E_{m + 1,\varphi} \right)^{*} - E_{m,z}\left( E_{m + 1,z} \right)^{*} -\right.\right.\\
    \left.\left.-iE_{m,\rho}\left( E_{m + 1,\varphi} \right)^{*} + iE_{m + 1,\rho}\left( E_{m,\varphi} \right)^{*} \right)n_{\rho} + 2E_{m,\rho}\left( E_{m,z} \right)^{*}n_{z} \right\rbrack
\end{multline}
\begin{multline}
    \left\langle{F}_y\right\rangle = \frac{\varepsilon_0}{4}\mathop{\mathrm{Re}}\int_{C_R}^{}{2\pi\rho\, \mathrm{d}c}\,\sum_{m = - \infty}^{+ \infty}\left\lbrack \left( E_{m,\rho}\left( E_{m + 1,\varphi} \right)^{*} + E_{m + 1,\rho}\left( E_{m,\varphi} \right)^{*} - iE_{m,\varphi}\left( E_{m + 1,\varphi} \right)^{*} +\right.\right.\\
    \left.\left.+ iE_{m,\rho}\left( E_{m + 1,\rho} \right)^{*} + iE_{m,z}\left( E_{m + 1,z} \right)^{*} \right)n_{\rho} + 2E_{m,\varphi}\left( E_{m,z} \right)^{*}n_{z} \right\rbrack
\end{multline}
\begin{multline}
    \left\langle{F}_{z}\right\rangle = \frac{\varepsilon_0}{4}\mathop{\mathrm{Re}}\int_{C_R}^{}{2\pi\rho\, \mathrm{d}c}\,\sum_{m = - \infty}^{+ \infty}\left\lbrack \left( E_{m,\rho}\left( E_{m + 1,\varphi} \right)^{*} + E_{m + 1,\rho}\left( E_{m,\varphi} \right)^{*} - iE_{m,\varphi}\left( E_{m + 1,\varphi} \right)^{*} +\right.\right.\\
    \left.\left.+ iE_{m,\rho}\left( E_{m + 1,\rho} \right)^{*} + iE_{m,z}\left( E_{m + 1,z} \right)^{*} \right)n_{\rho} + 2E_{m,\varphi}\left( E_{m,z} \right)^{*}n_{z} \right\rbrack
\end{multline}

We note that $n_\rho$ and $n_z$ in the expressions above are the components of the surface normal vector and thus may depend on $\rho$ and $z$.

\section{S3. Absorption, extinction, and scattering cross-section}

\subsection{Scattering cross-section}
The time-averaged Poyting vector for the scattered field can be written as
\begin{equation}
\mathbf{S}^\text{scat}=\frac{1}{2}\text{Re}\left\{\mathbf{E}\times\mathbf{H}^*\right\}=\frac{1}{2}\sum_m\sum_{m'}\text{Re}\left\{\mathbf{E}^{m}\times\mathbf{H}^{m'*}e^{i(m'-m)\varphi}\right\}    
\end{equation}
Therefore the total time-averaged scattered power $P^\text{scat}$ can be written as
\begin{align}
\label{eq:supp_scat1}
    & P^\text{scat}=\int_{S^2}\mathbf{S}^\text{scat} \mathbf{n} \dd s = \frac{1}{2}\sum_m\sum_{m'}\text{Re}\left\{\underbrace{\int_0^{2\pi} e^{i(m'-m)\varphi}\mathrm{d}\varphi}_{2\pi\delta_{mm'}}\int_{C_R} \mathbf{n}\cdot[\mathbf{E}^{m}\times\mathbf{H}^{m'*}]\ \rho \dd c\right\}     = \\
\label{eq:supp_scat2}
    & = \sum_m  \frac{1}{2}\text{Re}\left\{ 
    \int_{C_R} \mathbf{n}\cdot[\mathbf{E}^{m}\times\mathbf{H}^{m'*}] \rho\dd  c\right\}=\sum_m \int_{S^2}\mathbf{S}_m^\text{scat} \mathbf{n} \dd s =\sum_m P^\text{scat}_m.
\end{align}
Therefore, one can see that each azimuthal harmonic is scattered independently. The scattering cross-section can be calculated by definition:
\begin{equation}
    \sigma^\text{scat}=\frac{P^\text{scat}}{I_\text{inc}}=\sum_m\sigma^\text{scat}_m.
\end{equation}

\subsection{Absorption cross-section}
The time-averaged power $P^\text{abs}$ absorbed by the scatterer can be calculated according to the Joule low:
\begin{equation}
    P^\text{abs}=\frac{1}{2}\int \text{Re}\left\{\mathbf{j}^*\mathbf{E}\right\}\mathrm{d}V=
    \frac{1}{2}\int \text{Re}\left\{-i\omega\mathbf{P}^*\mathbf{E}\right\}\mathrm{d}V=
    \frac{\omega}{2}\int \text{Im}\left\{\mathbf{P}^*\mathbf{E}\right\}\mathrm{d}V.
\end{equation}
For the axially symmetric scatterer, one can rewrite it as  
\begin{align}
\label{eq:supp_abs1}
  &  P^\text{abs}=\frac{\omega}{2}\int \text{Im}\left\{\mathbf{P}^*\mathbf{E}\right\}\mathrm{d}V=\frac{\omega}{2}\text{Im}\left\{\ \sum_m\sum_{m'}\underbrace{\int_0^{2\pi} e^{i(m'-m)\varphi}\mathrm{d}\varphi}_{2\pi\delta_{mm'}}\iint_\Omega\mathbf{P}_m^*\mathbf{E}_{m'}\rho\mathrm{d}\rho\mathrm{d}z\right\}= \\
\label{eq:supp_abs2}
  & =\sum_m\pi\omega \iint_\Omega\text{Im}\left\{ \mathbf{P}_m^*\mathbf{E}_{m}\right\}\rho\mathrm{d}\rho\mathrm{d}z=\sum_m P^\text{abs}_m.
\end{align}
Therefore, one can see that each azimuthal harmonic is absorbed independently. The absorption cross-section can be calculated by definition:
\begin{equation}
    \sigma^\text{abs}=\frac{P^\text{abs}}{I_\text{inc}}=\sum_m\sigma^\text{abs}_m.
\end{equation}

\subsection{Extinction cross-section}
The total time-averaged power $P^\text{ext}$ taken by the scatterer can from the incident field be calculated according using the Lorentz reciprocity theorem~\cite{Snyder}. For the case of a finite size scatters one can write:
\begin{equation}
\label{eq:supp_reciprocity}
    -\nabla \left(
    \mathbf{E}^\text{inc}\times\mathbf{H}^{\text{scat}*}+    \mathbf{E}^{\text{scat}*}\times\mathbf{H}^{\text{inc}}\right)=\mathbf{j}^*\mathbf{E}^\text{inc}.
\end{equation}
Here $\mathbf{j}$ is the total currents inside the scatterer induced by the incident field. This relation allows for writing the expression for the extinction cross-section through the volume or surface integral, or thought the surface or linear integral in the reduced 2D geometry. The $P^\text{ext}$ can be written as:
\begin{equation}
    P^\text{ext}=\frac{1}{2}\int \text{Re}\left\{\mathbf{j}^*\mathbf{E}^\text{inc}\right\}\mathrm{d}V=
    \frac{1}{2}\int \text{Re}\left\{-i\omega\mathbf{P}^*\mathbf{E}^\text{inc}\right\}\mathrm{d}V=
    \frac{\omega}{2}\int \text{Im}\left\{\mathbf{P}^*\mathbf{E}^\text{inc}\right\}\mathrm{d}V.
\end{equation}
By analogy with Eqs.~\eqref{eq:supp_abs1} and\eqref{eq:supp_abs2} one can write
\begin{equation}
     P^\text{ext}=\sum_m\pi\omega \iint_\Omega\text{Im}\left\{\ \mathbf{P}_m^*\mathbf{E}^\text{inc}_{m}\right\}\rho\mathrm{d}\rho\mathrm{d}z=\sum_m P^\text{ext}_m.
\end{equation}
At the same time, using Eq.~\eqref{eq:supp_reciprocity} the extincted power can be written as
\begin{align}
P^\text{ext}=-\frac{1}{2}\int_{S^2}\text{Re}\left(
    \mathbf{E}^\text{inc}\times\mathbf{H}^{\text{scat}*}+    \mathbf{E}^{\text{scat}*}\times\mathbf{H}^{\text{inc}}\right) \mathbf{n} \dd s.
\end{align}
By analogy with Eqs.~\eqref{eq:supp_scat1} and \eqref{eq:supp_scat2}, one can simplify this equation to the following form:
\begin{align}
   P^\text{ext} = -\pi\sum_m\int_{C_R} \text{Re}\left(
    \mathbf{E}_m^\text{inc}\times\mathbf{H}_m^{\text{scat}*}+    \mathbf{E}_m^{\text{scat}*}\times\mathbf{H}_m^{\text{inc}}\right) \mathbf{n} \rho \dd c = \sum_m P^\text{ext}_m
\end{align}

\section{S4. Components of the second order nonlinear susceptibility tensor}

\begin{widetext}
All components of the second order nonlinear susceptibility tensor $\vb{e}_{i}\vb{e}_{j}\vb{e}_{k}$ in a cylindrical coordinate system:
\begin{equation}
\vb{e}_{x}\vb{e}_{y}\vb{e}_{z}=
\frac{e^{2i\varphi}-e^{-2i\varphi}}{4i}\vb{e}_{\rho}\vb{e}_{\rho}\vb{e}_{z}-
\frac{e^{2i\varphi}-e^{-2i\varphi}}{4i}\vb{e}_{\varphi}\vb{e}_{\varphi}\vb{e}_{z}+
\frac{e^{2i\varphi}+e^{-2i\varphi}-2}{4}\vb{e}_{\varphi}\vb{e}_{\rho}\vb{e}_{z}+
\frac{e^{2i\varphi}+e^{-2i\varphi}+2}{4}\vb{e}_{\rho}\vb{e}_{\varphi}\vb{e}_{z}
\end{equation}
\begin{equation}
\vb{e}_{y}\vb{e}_{x}\vb{e}_{z}=
\frac{e^{2i\varphi}-e^{-2i\varphi}}{4i}\vb{e}_{\rho}\vb{e}_{\rho}\vb{e}_{z}-
\frac{e^{2i\varphi}-e^{-2i\varphi}}{4i}\vb{e}_{\varphi}\vb{e}_{\varphi}\vb{e}_{z}+
\frac{e^{2i\varphi}+e^{-2i\varphi}-2}{4}\vb{e}_{\rho}\vb{e}_{\varphi}\vb{e}_{z}+
\frac{e^{2i\varphi}+e^{-2i\varphi}+2}{4}\vb{e}_{\varphi}\vb{e}_{\rho}\vb{e}_{z}
\end{equation}

{\begin{equation}
\begin{aligned}
\vb{e}_{x}\vb{e}_{y}\vb{e}_{x}=\\={e^{3i\varphi}}\left(\frac1{8i}\vb e_{\rho}\vb e_{\rho}\vb e_{\rho}-\frac1{8i}(\vb e_{\varphi}\vb e_{\varphi}\vb e_{\rho}+\vb e_{\rho}\vb e_{\varphi}\vb e_{\varphi}+\vb e_{\varphi}\vb e_{\rho}\vb e_{\varphi})+\frac1{8}(\vb e_{\rho}\vb e_{\varphi}\vb e_{\rho}+\vb e_{\rho}\vb e_{\rho}\vb e_{\varphi}+\vb e_{\varphi}\vb e_{\rho}\vb e_{\rho})-\frac1{8}\vb e_{\varphi}\vb e_{\varphi}\vb e_{\varphi}\right)+\\+
{e^{-3i\varphi}}\left(-\frac1{8i}\vb e_{\rho}\vb e_{\rho}\vb e_{\rho}+\frac1{8i}(\vb e_{\varphi}\vb e_{\varphi}\vb e_{\rho}+\vb e_{\rho}\vb e_{\varphi}\vb e_{\varphi}+\vb e_{\varphi}\vb e_{\rho}\vb e_{\varphi})+\frac1{8}(\vb e_{\rho}\vb e_{\varphi}\vb e_{\rho}+\vb e_{\rho}\vb e_{\rho}\vb e_{\varphi}+\vb e_{\varphi}\vb e_{\rho}\vb e_{\rho})-\frac1{8}\vb e_{\varphi}\vb e_{\varphi}\vb e_{\varphi}\right)+\\+
{e^{i\varphi}}\left(\frac1{8i}\vb e_{\rho}\vb e_{\rho}\vb e_{\rho}+\frac1{8i}(3\vb e_{\varphi}\vb e_{\rho}\vb e_{\varphi}-\vb e_{\varphi}\vb e_{\varphi}\vb e_{\rho}-\vb e_{\rho}\vb e_{\varphi}\vb e_{\varphi})+\frac1{8}(3\vb e_{\rho}\vb e_{\varphi}\vb e_{\rho}-\vb e_{\rho}\vb e_{\rho}\vb e_{\varphi}-\vb e_{\varphi}\vb e_{\rho}\vb e_{\rho})+\frac1{8}\vb e_{\varphi}\vb e_{\varphi}\vb e_{\varphi}\right)+\\+
{e^{-i\varphi}}\left(-\frac1{8i}\vb e_{\rho}\vb e_{\rho}\vb e_{\rho}-\frac1{8i}(3\vb e_{\varphi}\vb e_{\rho}\vb e_{\varphi}-\vb e_{\varphi}\vb e_{\varphi}\vb e_{\rho}-\vb e_{\rho}\vb e_{\varphi}\vb e_{\varphi})+\frac1{8}(3\vb e_{\rho}\vb e_{\varphi}\vb e_{\rho}-\vb e_{\rho}\vb e_{\rho}\vb e_{\varphi}-\vb e_{\varphi}\vb e_{\rho}\vb e_{\rho})+\frac1{8}\vb e_{\varphi}\vb e_{\varphi}\vb e_{\varphi}\right) \ \ \ 
\end{aligned}
\end{equation}}

{\begin{equation}
\begin{aligned}
\vb{e}_{x}\vb{e}_{y}\vb{e}_{y}=\\={e^{3i\varphi}}\left(-\frac1{8}\vb e_{\rho}\vb e_{\rho}\vb e_{\rho}+\frac1{8}(\vb e_{\varphi}\vb e_{\varphi}\vb e_{\rho}+\vb e_{\rho}\vb e_{\varphi}\vb e_{\varphi}+\vb e_{\varphi}\vb e_{\rho}\vb e_{\varphi})+\frac1{8i}(\vb e_{\rho}\vb e_{\varphi}\vb e_{\rho}+\vb e_{\rho}\vb e_{\rho}\vb e_{\varphi}+\vb e_{\varphi}\vb e_{\rho}\vb e_{\rho})-\frac1{8i}\vb e_{\varphi}\vb e_{\varphi}\vb e_{\varphi}\right)+\\+
{e^{-3i\varphi}}\left(-\frac1{8}\vb e_{\rho}\vb e_{\rho}\vb e_{\rho}+\frac1{8}(\vb e_{\varphi}\vb e_{\varphi}\vb e_{\rho}+\vb e_{\rho}\vb e_{\varphi}\vb e_{\varphi}+\vb e_{\varphi}\vb e_{\rho}\vb e_{\varphi})-\frac1{8i}(\vb e_{\rho}\vb e_{\varphi}\vb e_{\rho}+\vb e_{\rho}\vb e_{\rho}\vb e_{\varphi}+\vb e_{\varphi}\vb e_{\rho}\vb e_{\rho})+\frac1{8i}\vb e_{\varphi}\vb e_{\varphi}\vb e_{\varphi}\right)+\\+
{e^{i\varphi}}\left(\frac1{8}\vb e_{\rho}\vb e_{\rho}\vb e_{\rho}-\frac1{8}(\vb e_{\varphi}\vb e_{\rho}\vb e_{\varphi}+\vb e_{\varphi}\vb e_{\varphi}\vb e_{\rho}-3\vb e_{\rho}\vb e_{\varphi}\vb e_{\varphi})+\frac1{8i}(\vb e_{\rho}\vb e_{\varphi}\vb e_{\rho}+\vb e_{\rho}\vb e_{\rho}\vb e_{\varphi}-3\vb e_{\varphi}\vb e_{\rho}\vb e_{\rho})-\frac1{8i}\vb e_{\varphi}\vb e_{\varphi}\vb e_{\varphi}\right)+\\+
{e^{-i\varphi}}\left(\frac1{8}\vb e_{\rho}\vb e_{\rho}\vb e_{\rho}-\frac1{8}(\vb e_{\varphi}\vb e_{\rho}\vb e_{\varphi}+\vb e_{\varphi}\vb e_{\varphi}\vb e_{\rho}-3\vb e_{\rho}\vb e_{\varphi}\vb e_{\varphi})+\frac1{8i}(-\vb e_{\rho}\vb e_{\varphi}\vb e_{\rho}-\vb e_{\rho}\vb e_{\rho}\vb e_{\varphi}+\vb 3e_{\varphi}\vb e_{\rho}\vb e_{\rho})+\frac1{8i}\vb e_{\varphi}\vb e_{\varphi}\vb e_{\varphi}\right) \ \ \ 
\end{aligned}
\end{equation}}

\begin{equation}
\begin{aligned}
\vb{e}_{x}\vb{e}_{x}\vb{e}_{z}=e^{0i\varphi}\left(\frac12\vb{e}_{\rho} \vb{e}_{\rho} \vb{e}_{z} +\frac12 \vb{e}_{\varphi} \vb{e}_{\varphi} \vb{e}_{z} \right) +\\+
e^{2i\varphi}\left(\frac14\vb{e}_{\rho} \vb{e}_{\rho} \vb{e}_{z} -\frac14 \vb{e}_{\varphi} \vb{e}_{\varphi} \vb{e}_{z} -\frac1{4i} (\vb{e}_{\varphi} \vb{e}_{\rho} \vb{e}_{z}+\vb{e}_{\rho} \vb{e}_{\varphi} \vb{e}_{z}) \right) +\\+
e^{-2i\varphi}\left(\frac14\vb{e}_{\rho} \vb{e}_{\rho} \vb{e}_{z} -\frac14 \vb{e}_{\varphi} \vb{e}_{\varphi} \vb{e}_{z} +\frac1{4i} (\vb{e}_{\varphi} \vb{e}_{\rho} \vb{e}_{z}+\vb{e}_{\rho} \vb{e}_{\varphi} \vb{e}_{z}) \right) \ \ \ 
\end{aligned}
\end{equation}

\begin{equation}
\begin{aligned}
\vb{e}_{y}\vb{e}_{y}\vb{e}_{z}=e^{0i\varphi}\left(\frac12\vb{e}_{\rho} \vb{e}_{\rho} \vb{e}_{z} +\frac12 \vb{e}_{\varphi} \vb{e}_{\varphi} \vb{e}_{z} \right) +\\+
e^{2i\varphi}\left(-\frac14\vb{e}_{\rho} \vb{e}_{\rho} \vb{e}_{z} +\frac14 \vb{e}_{\varphi} \vb{e}_{\varphi} \vb{e}_{z} +\frac1{4i} (\vb{e}_{\varphi} \vb{e}_{\rho} \vb{e}_{z}+\vb{e}_{\rho} \vb{e}_{\varphi} \vb{e}_{z}) \right) +\\+
e^{-2i\varphi}\left(-\frac14\vb{e}_{\rho} \vb{e}_{\rho} \vb{e}_{z} +\frac14 \vb{e}_{\varphi} \vb{e}_{\varphi} \vb{e}_{z} -\frac1{4i} (\vb{e}_{\varphi} \vb{e}_{\rho} \vb{e}_{z}+\vb{e}_{\rho} \vb{e}_{\varphi} \vb{e}_{z}) \right) \ \ \ 
\end{aligned}
\end{equation}

\begin{equation}
\begin{aligned}
\vb{e}_{x}\vb{e}_{x}\vb{e}_{z}+\vb{e}_{y}\vb{e}_{y}\vb{e}_{z}=e^{0i\varphi}\left(\vb{e}_{\rho} \vb{e}_{\rho} \vb{e}_{z} +  \vb{e}_{\varphi} \vb{e}_{\varphi} \vb{e}_{z} \right)
\end{aligned}
\end{equation}

\begin{equation}
\begin{aligned}
\vb{e}_{x}\vb{e}_{x}\vb{e}_{x}= \\
= {e^{3i\varphi}}\left(\frac18\vb e_{\rho}\vb e_{\rho}\vb e_{\rho}-\frac18(\vb e_{\varphi}\vb e_{\varphi}\vb e_{\rho}+\vb e_{\rho}\vb e_{\varphi}\vb e_{\varphi}+\vb e_{\varphi}\vb e_{\rho}\vb e_{\varphi})-\frac1{8i}(\vb e_{\rho}\vb e_{\varphi}\vb e_{\rho}+\vb e_{\rho}\vb e_{\rho}\vb e_{\varphi}+\vb e_{\varphi}\vb e_{\rho}\vb e_{\rho})+\frac1{8i}\vb e_{\varphi}\vb e_{\varphi}\vb e_{\varphi}\right)+\\+
{e^{-3i\varphi}}\left(\frac18\vb e_{\rho}\vb e_{\rho}\vb e_{\rho}-\frac18(\vb e_{\varphi}\vb e_{\varphi}\vb e_{\rho}+\vb e_{\rho}\vb e_{\varphi}\vb e_{\varphi}+\vb e_{\varphi}\vb e_{\rho}\vb e_{\varphi})+\frac1{8i}(\vb e_{\rho}\vb e_{\varphi}\vb e_{\rho}+\vb e_{\rho}\vb e_{\rho}\vb e_{\varphi}+\vb e_{\varphi}\vb e_{\rho}\vb e_{\rho})-\frac1{8i}\vb e_{\varphi}\vb e_{\varphi}\vb e_{\varphi}\right)+\\+
{e^{i\varphi}}\left(\frac38\vb e_{\rho}\vb e_{\rho}\vb e_{\rho}+\frac18(\vb e_{\varphi}\vb e_{\varphi}\vb e_{\rho}+\vb e_{\rho}\vb e_{\varphi}\vb e_{\varphi}+\vb e_{\varphi}\vb e_{\rho}\vb e_{\varphi})-\frac1{8i}(\vb e_{\rho}\vb e_{\varphi}\vb e_{\rho}+\vb e_{\rho}\vb e_{\rho}\vb e_{\varphi}+\vb e_{\varphi}\vb e_{\rho}\vb e_{\rho})-\frac3{8i}\vb e_{\varphi}\vb e_{\varphi}\vb e_{\varphi}\right)+\\+
{e^{-i\varphi}}\left(\frac38\vb e_{\rho}\vb e_{\rho}\vb e_{\rho}+\frac18(\vb e_{\varphi}\vb e_{\varphi}\vb e_{\rho}+\vb e_{\rho}\vb e_{\varphi}\vb e_{\varphi}+\vb e_{\varphi}\vb e_{\rho}\vb e_{\varphi})+\frac1{8i}(\vb e_{\rho}\vb e_{\varphi}\vb e_{\rho}+\vb e_{\rho}\vb e_{\rho}\vb e_{\varphi}+\vb e_{\varphi}\vb e_{\rho}\vb e_{\rho})+\frac3{8i}\vb e_{\varphi}\vb e_{\varphi}\vb e_{\varphi}\right) \ \ \ 
\end{aligned}
\end{equation}

\begin{equation}
\vb{e}_{x}\vb{e}_{z}\vb{e}_{z}={e^{i\varphi}}\left(\frac1{2}\vb{e}_{\rho}\vb{e}_{z}\vb{e}_{z}-\frac1{2i}\vb{e}_{\varphi}\vb{e}_{z}\vb{e}_{z}\right)+
{e^{-i\varphi}}\left(\frac1{2}\vb{e}_{\rho}\vb{e}_{z}\vb{e}_{z}+\frac1{2i}\vb{e}_{\varphi}\vb{e}_{z}\vb{e}_{z}\right)
\end{equation}

\begin{equation}
\vb{e}_{y}\vb{e}_{z}\vb{e}_{z}={e^{i\varphi}}\left(\frac1{2i}\vb{e}_{\rho}\vb{e}_{z}\vb{e}_{z}+\frac1{2}\vb{e}_{\varphi}\vb{e}_{z}\vb{e}_{z}\right)+
{e^{-i\varphi}}\left(-\frac1{2i}\vb{e}_{\rho}\vb{e}_{z}\vb{e}_{z}+\frac1{2}\vb{e}_{\varphi}\vb{e}_{z}\vb{e}_{z}\right)
\end{equation}

{\begin{equation}
\begin{aligned}
\vb{e}_{y}\vb{e}_{y}\vb{e}_{y}=\\= {e^{3i\varphi}}\left(-\frac1{8i}\vb e_{\rho}\vb e_{\rho}\vb e_{\rho}+\frac1{8i}(\vb e_{\varphi}\vb e_{\varphi}\vb e_{\rho}+\vb e_{\rho}\vb e_{\varphi}\vb e_{\varphi}+\vb e_{\varphi}\vb e_{\rho}\vb e_{\varphi})-\frac1{8}(\vb e_{\rho}\vb e_{\varphi}\vb e_{\rho}+\vb e_{\rho}\vb e_{\rho}\vb e_{\varphi}+\vb e_{\varphi}\vb e_{\rho}\vb e_{\rho})+\frac1{8}\vb e_{\varphi}\vb e_{\varphi}\vb e_{\varphi}\right)+\\+
{e^{-3i\varphi}}\left(\frac1{8i}\vb e_{\rho}\vb e_{\rho}\vb e_{\rho}-\frac1{8i}(\vb e_{\varphi}\vb e_{\varphi}\vb e_{\rho}+\vb e_{\rho}\vb e_{\varphi}\vb e_{\varphi}+\vb e_{\varphi}\vb e_{\rho}\vb e_{\varphi})-\frac1{8}(\vb e_{\rho}\vb e_{\varphi}\vb e_{\rho}+\vb e_{\rho}\vb e_{\rho}\vb e_{\varphi}+\vb e_{\varphi}\vb e_{\rho}\vb e_{\rho})+\frac1{8}\vb e_{\varphi}\vb e_{\varphi}\vb e_{\varphi}\right)+\\+
{e^{i\varphi}}\left(\frac3{8i}\vb e_{\rho}\vb e_{\rho}\vb e_{\rho}+\frac1{8i}(\vb e_{\varphi}\vb e_{\varphi}\vb e_{\rho}+\vb e_{\rho}\vb e_{\varphi}\vb e_{\varphi}+\vb e_{\varphi}\vb e_{\rho}\vb e_{\varphi})+\frac1{8}(\vb e_{\rho}\vb e_{\varphi}\vb e_{\rho}+\vb e_{\rho}\vb e_{\rho}\vb e_{\varphi}+\vb e_{\varphi}\vb e_{\rho}\vb e_{\rho})+\frac3{8}\vb e_{\varphi}\vb e_{\varphi}\vb e_{\varphi}\right)+\\+
{e^{-i\varphi}}\left(-\frac3{8i}\vb e_{\rho}\vb e_{\rho}\vb e_{\rho}-\frac1{8i}(\vb e_{\varphi}\vb e_{\varphi}\vb e_{\rho}+\vb e_{\rho}\vb e_{\varphi}\vb e_{\varphi}+\vb e_{\varphi}\vb e_{\rho}\vb e_{\varphi})+\frac1{8}(\vb e_{\rho}\vb e_{\varphi}\vb e_{\rho}+\vb e_{\rho}\vb e_{\rho}\vb e_{\varphi}+\vb e_{\varphi}\vb e_{\rho}\vb e_{\rho})+\frac3{8}\vb e_{\varphi}\vb e_{\varphi}\vb e_{\varphi}\right) \ \ \ 
\end{aligned}
\end{equation}}

\begin{equation}
\vb{e}_{z}\vb{e}_{z}\vb{e}_{z}= \vb{e}_{z}\vb{e}_{z}\vb{e}_{z}
\end{equation}

\begin{equation}
\begin{aligned}
\vb{e}_{y}\vb{e}_{x}\vb{e}_{z}+\vb{e}_{x}\vb{e}_{y}\vb{e}_{z}= \\ =
{e^{2i\varphi}}\left(\frac1{2i}\vb{e}_{\rho}\vb{e}_{\rho}\vb{e}_{z}-\frac1{2i}\vb{e}_{\varphi}\vb{e}_{\varphi}\vb{e}_{z}+\frac1{2}\vb{e}_{\varphi}\vb{e}_{\rho}\vb{e}_{z}+\frac1{2}\vb{e}_{\rho}\vb{e}_{\varphi}\vb{e}_{z}\right)+ \\+
{e^{-2i\varphi}}\left(-\frac1{2i}\vb{e}_{\rho}\vb{e}_{\rho}\vb{e}_{z}+\frac1{2i}\vb{e}_{\varphi}\vb{e}_{\varphi}\vb{e}_{z}+\frac1{2}\vb{e}_{\varphi}\vb{e}_{\rho}\vb{e}_{z}+\frac1{2}\vb{e}_{\rho}\vb{e}_{\varphi}\vb{e}_{z}\right)
\end{aligned}
\end{equation}

\begin{equation}
\begin{aligned}
\vb{e}_{y}\vb{e}_{z}\vb{e}_{x}+\vb{e}_{x}\vb{e}_{z}\vb{e}_{y}= \\ =
{e^{2i\varphi}}\left(\frac1{2i}\vb{e}_{\rho}\vb{e}_{z}\vb{e}_{\rho}-\frac1{2i}\vb{e}_{\varphi}\vb{e}_{z}\vb{e}_{\varphi}+\frac1{2}\vb{e}_{\varphi}\vb{e}_{z}\vb{e}_{\rho}+\frac1{2}\vb{e}_{\rho}\vb{e}_{z}\vb{e}_{\varphi}\right)+ \\+
{e^{-2i\varphi}}\left(-\frac1{2i}\vb{e}_{\rho}\vb{e}_{z}\vb{e}_{\rho}+\frac1{2i}\vb{e}_{\varphi}\vb{e}_{z}\vb{e}_{\varphi}+\frac1{2}\vb{e}_{\varphi}\vb{e}_{z}\vb{e}_{\rho}+\frac1{2}\vb{e}_{\rho}\vb{e}_{z}\vb{e}_{\varphi}\right)
\end{aligned}
\end{equation}

\begin{equation}
\begin{aligned}
\vb{e}_{z}\vb{e}_{y}\vb{e}_{x}+\vb{e}_{z}\vb{e}_{x}\vb{e}_{y}= \\ =
{e^{2i\varphi}}\left(\frac1{2i}\vb{e}_{z}\vb{e}_{\rho}\vb{e}_{\rho}-\frac1{2i}\vb{e}_{z}\vb{e}_{\varphi}\vb{e}_{\varphi}+\frac1{2}\vb{e}_{z}\vb{e}_{\varphi}\vb{e}_{\rho}+\frac1{2}\vb{e}_{z}\vb{e}_{\rho}\vb{e}_{\varphi}\right)+ \\+
{e^{-2i\varphi}}\left(-\frac1{2i}\vb{e}_{z}\vb{e}_{\rho}\vb{e}_{\rho}+\frac1{2i}\vb{e}_{z}\vb{e}_{\varphi}\vb{e}_{\varphi}+\frac1{2}\vb{e}_{z}\vb{e}_{\varphi}\vb{e}_{\rho}+\frac1{2}\vb{e}_{z}\vb{e}_{\rho}\vb{e}_{\varphi}\right)
\end{aligned}
\end{equation}
\end{widetext}

\bibliography{Artikel}